%% file: Sources_Estimation.tex
\documentclass[10pt, twocolumn, final]{IEEEtran}%
\input{preamble.tex}

\usepackage{caption}
\newcommand{\Exp}{\text{Exp}}
\newcommand{\ER}{\text{ER}}
\newcommand{\BA}{\text{BA}}

\graphicspath{{Figures/},{MSEG_figures/}} 
\usepackage{makecell}
\newenvironment{bsmallmatrix}{\left[\begin{smallmatrix}}	{\end{smallmatrix}\right]}
\newsavebox\smlmat
\savebox\smlmat{$\begin{bsmallmatrix} 
	4 & 3 & 0 \\
	3 & 4 & 0 \\
	0 & 0 & 4 
	\end{bsmallmatrix}$}

\newsavebox\sml
\savebox\sml{$\begin{bsmallmatrix} 
	4 & 2 & 2 \\
	2 & 4 & 2 \\
	2 & 2 & 4 
	\end{bsmallmatrix}$}

\begin{document}
\title{Estimating Infection Sources in Networks Using Partial Timestamps}
\author{Wenchang Tang, Feng Ji, and Wee Peng Tay,~\IEEEmembership{Senior Member,~IEEE}
    \thanks{This work was supported in part by the Singapore Ministry of Education Academic Research Fund Tier 2 grant MOE2014-T2-1-028.}%
	\thanks{The authors are with the School of Electrical and Electronic Engineering, Nanyang Technological University, Singapore (e-mail: E150012@e.ntu.edu.sg, jifeng@ntu.edu.sg, wptay@ntu.edu.sg). }
}
\maketitle

\begin{abstract}
We study the problem of identifying infection sources in a network based on the network topology, and a subset of infection timestamps. In the case of a single infection source in a tree network, we derive the maximum likelihood estimator of the source and the unknown diffusion parameters. We then introduce a new heuristic involving an optimization over a parametrized family of Gromov matrices to develop a single source estimation algorithm for general graphs. Compared with the breadth-first search tree heuristic commonly adopted in the literature, simulations demonstrate that our approach achieves better estimation accuracy than several other benchmark algorithms, even though these require more information like the diffusion parameters. We next develop a multiple sources estimation algorithm for general graphs, which first partitions the graph into source candidate clusters, and then applies our single source estimation algorithm to each cluster. We show that if the graph is a tree, then each source candidate cluster contains at least one source. Simulations using synthetic and real networks, and experiments using real-world data suggest that our proposed algorithms are able to estimate the true infection source(s) to within a small number of hops with a small portion of the infection timestamps being observed.
\end{abstract}


\section{Introduction}\label{sect:intro}

Online social networks such as Facebook, Twitter and Sina Weibo have grown immensely in recent decades. A rumor or piece of news can be shared and forwarded with high speed through friendship circles \cite{Kumar2006}. Similar propagation phenomena in complex networks are ubiquitous in nature and modern society. For example, viruses propagate throughout the Internet and infect millions of computers or mobile phones\cite{Lloyd2001, Wang1071}. Disease epidemics can quickly spread in human society \cite{Neumann2009,Pastor2001}. We regard such rumors, viruses and epidemics as infections, and the propagation initiated by an infection source can be modeled using a diffusion or infection process in networks\cite{Guille2013,Pastor2001}. Such infections may incur massive losses to society \cite{berghel2001code,richard2008}. In view of this, it is often important to be able to accurately and promptly identify the infection sources, so that proper control measures can be adopted. In this paper, we consider the infection source estimation problem based on knowledge of the network topology, and a subset of infection timestamps.

\subsection{Related Works}\label{subsec:related}
Several works have tackled the problem of infection source estimation under different assumptions. Many of these works are based on the knowledge of network topology and the infection status of a portion of nodes in the network. In the literature, various spreading models have been proposed including the Susceptible-Infected (SI) \cite{Shah2010}, Susceptible-Infected-Recovered (SIR) \cite{Zhu2013}, and Susceptible-Infected-Susceptible (SIS) \cite{LuoTay2013} models. Specifically, in the SI model, an infected node remains infected forever; in the SIR model, it can recover and cannot be further infected; and in the SIS model, a recovered node can become infected again. A rumor centrality estimator under the SI model was developed in \cite{Shah2010, Shah2012}, while \cite{Luo2013, Fioriti2014, Prakash2012,Ji2017} developed estimators for identifying multiple infection sources under the SI model. The paper \cite{Luo2014} considers infection source estimation when only a subset of infected nodes are observed. In \cite{Wang2014}, multiple observations of a SI spreading process are used for source estimation. In \cite{Zhu2013} and \cite{Zhu2016}, the problem of infection source estimation is investigated under the SIR model, and \cite{LuoTay2013} considers the SIS model. All the aforementioned works perform source estimation based on the observed status of the nodes, and the network topology. Other related works assume additionally a priori knowledge like the infection spreading rate, including \cite{Altarelli2014} which performs source inference via belief propagation, and \cite{Lokhov2014} which developed a dynamic message passing algorithm. Recently, \cite{zhu2015source} considered the alternative Independent Cascade (IC) model for infection spreading, and proposed the Short-Fat Tree (SFT) algorithm for source identification under the IC model. The reference \cite{hu2017} developed a framework for optimal source estimation in arbitrary weighted networks with arbitrary distribution of sources by combining controllability theory and compressive sensing. The reference \cite{ Kumar2017} considered the case where additional relative information about the infection times of a fraction of node pairs is also available. Finally, the works \cite{Luo2015,Fanti2015} have studied the complementary problem of source obfuscation and proposed messaging protocols which can spread the infection as widely as possible while protecting the anonymity of the source.

Infection source estimation has also been investigated using infection timestamps. In this framework, we make use of observations of the first infection times of a subset of nodes in the network. In \cite{Pinto2012}, an algorithm based on the maximum likelihood estimator (MLE) is proposed. Two ranking algorithms using a modified breadth-first search (BFS) tree heuristic for general graphs have been proposed in \cite{Zhu2014}, which demonstrate improved estimation accuracy compared to the algorithm in \cite{Pinto2012}. In \cite{Louni2014}, a two-stage algorithm was proposed to locate a single source in large networks. The reference \cite{TanTay:C17} proposed a sequential source estimation algorithm that allows online update of the source estimate as timestamps are observed sequentially. In \cite{Shen2016}, a time-reversal backward spreading (TRBS) algorithm was proposed to infer a single source in a weighted network. The papers \cite{ zejn2013}, \cite{ zejn2015 } discuss the selection of observer nodes under the deterministic slotted SI model, so as to achieve low probability of error detection. The problem of multiple sources estimation with infection timestamps is less studied. In \cite{fu2016multi}, a backward diffusion based method was proposed. However, the paper assumes that all infection sources initiate diffusion at the same start time, and the mean of the infection propagation delays along all edges and the number of sources are known. Such assumptions can be restrictive for practical applications. 

\subsection{Our Contributions}
We first study the single source estimation problem with less assumptions compared with the works \cite{Pinto2012, Louni2014, TanTay:C17}. Similar to \cite{Pinto2012, Louni2014, TanTay:C17}, we assume that the propagation delay along each edge can be modeled using a Gaussian distribution. However, unlike these references, the diffusion parameters, including the mean and variance of the infection propagation delay, are assumed to be unknown. When the graph is a tree, we derive the optimal MLE that simultaneously estimates the source and diffusion parameters. It is common in the literature \cite{Shah2010, zheng2015, Pinto2012, Louni2014} to use a breadth-first search (BFS) tree  heuristic to generalize an algorithm for identifying a single source in a tree to one for a general graph: for each node in the general graph, one constructs a BFS tree and assumes that any infection starting at this node diffuses along this BFS tree. In this paper, we introduce a novel heuristic, which involves finding an optimal convex combination of two BFS trees and a target covariance matrix (cf. \Cref{subsect:GSSI}) for each node to approximate the diffusion process better, before generalizing our algorithm for trees to general graphs. We call this the Gromov Single Source Identification (GSSI) algorithm. Simulation results demonstrate that GSSI performs better than algorithms proposed in \cite{Pinto2012, Shen2016} although GSSI requires less information. 

We also study the case of multiple sources and propose a Source Candidate Clustering and Estimation (SCCE) algorithm, which makes less assumptions compared with the work \cite{fu2016multi} as we do not assume that infection sources have the same diffusion parameters, and the number of sources is also unknown a priori. SCCE can be divided into two steps. In the first step, we reduce the multiple sources estimation problem to the single source estimation problem by partitioning the graph into source candidate clusters. We show that if the graph is a tree, then each cluster contains at least one source. We then apply a procedure to estimate the number of sources in each cluster and estimate the infection region of each source. In the second step, we apply the GSSI algorithm in each region.  Simulation results demonstrate that our approach performs better than the algorithm proposed in \cite{fu2016multi}, and is able to accurately infer multiple sources with a small portion of timestamps.

The rest of this paper is organized as follows. In \cref{sect:formulation}, we formulate the infection source estimation problem and provide an interpretation based on Gromov matrices. In \cref{sect:single source}, we first derive the MLE for estimating a single source in a tree, and then present our GSSI algorithm using Gromov matrices for identifying a source in a general graph. Simulations on both synthetic and real-world networks are also provided to compare the performance of GSSI with various other algorithms in the literature. In \cref{sect:multiple sources}, we investigate the problem of multiple sources estimation, present our SCCE algorithm, and provide simulation results to verify its performance. \Cref{sect:conclusion} concludes the paper. 

\emph{Notations:} We use $\calN (\mu,\sigma^2)$ to denote the Gaussian distribution with mean $\mu$ and variance $\sigma^2$, and $\calT\calN(\mu,\sigma^2)$ to denote the corresponding truncated Gaussian distribution in which the distribution support is constrained to be non-negative. We use $\Exp(\lambda)$ to denote the exponential distribution with rate $\lambda$. For a set $X$, its cardinality is given by $|X|$. Given a matrix $\bA$, we use $[\bA]_{i,j}$ to denote its $(i,j)$-th entry. We write $\bA'$ for the transpose of $\bA$, $\bA^{-1}$ for its inverse, $\tr(\bA)$ for its trace, and $\det \bA$ for its determinant. For two nodes $u$ and $v$ in a tree, we let $[u,v]$ denote the path from $u$ to $v$, $(u,v]$ the path with the end node $u$ excluded, and so on. $\E[\cdot]$ is the expectation operator. The symbol $\sim$ means equality in distribution. 

\section{Problem Formulation }\label{sect:formulation}
We model the network as an undirected graph $G = (V,E)$, where $V$ is the set of nodes and $E$ is the set of edges. Suppose a collection of infection sources $S=\{s_1,...,s_{|S|}\}\subset V$ initiate a diffusion process at unknown start times on $G$. The infection propagates from an infected node to its neighbors stochastically along the edges connecting them. Let $\tau_{uv}$ be the random propagation delay associated with the edge connecting nodes $u$ and $v$. For example, if $u$ is first infected at time $t_u$, then the infection propagating from $u$ reaches $v$ at time $t_u+\tau_{uv}$. Similar to \cite{Pinto2012, Louni2014, TanTay:C17}, we assume that the propagation delays $\{\tau_{uv}\}$ for all pairs of adjacent nodes $u,v$ are independent and identically distributed (i.i.d.) continuous random variables that follow a Gaussian distribution $\calN (\mu,\sigma^2)$.\footnote{The Gaussian distribution is adopted for technical convenience although its realizations are not necessarily positive. In our simulations, we will generate the propagation delays using a truncated Gaussian distribution instead.} However, we also assume that the parameters $(\mu,\sigma^2)$ are unknown, and need to be estimated.

We assume the diffusion follows the SI model: any infected node never recovers. Suppose that we observe a collection of infected nodes $\calV=\{v_1,\ldots,v_n\}$, which is a subset of all the infected nodes, and the vector of their associated infection timestamps $\bT=[t_1,...,t_n]'$. Our goal is to estimate $S$ using $(\calV, \bT)$. To simplify our exposition and avoid non-invertible matrices in our cost functions in \cref{eq:ml_pro,eq:sse_para,eq:sse_para_graph}, we assume that $S \cap \calV = \emptyset$. There is no loss of generality making this assumption since one can generalize our methods by adding an additional step in which each vertex in $\calV$ is considered a potential source candidate, computing an appropriate variant of the respective cost functions in \cref{eq:ml_pro,eq:sse_para,eq:sse_para_graph} for these potential source nodes, and then choosing the node from amongst $\calV$ and that estimated by our method with the lowest cost to be the final estimate.

In this paper, we consider both the single source ($|S|=1$) and multiple sources ($|S|>1$) estimation problems. In the case where $|S|>1$, we assume that $|S|$ is unknown. 

\subsection{A Gromov Matrix Interpretation}\label{subsec:reformulation}

For two vertices $u,v$ in a tree $T \subset G$, let $d_T(u,v)$ be the length of the unique path between $u$ and $v$ (assuming each edge has length $1$). For a node $s$ of $T$, the \emph{Gromov product} (Definition 2.6 of \cite{MR1921706}) of $u$ and $v$ in $T$ with respect to (w.r.t.) $s$ is given by 
\begin{align}
(u,v)_{s} = \frac{1}{2}(d_T(u,s)+d_T(v,s)-d_T(u,v)).
\end{align}
The following lemma summarizes some simple properties of the Gromov product.

\begin{Lemma}\label{lem:Gromov}
Let $s$, $u$ and $v$ be nodes of a tree $T$.
\begin{enumerate}[(i)]
\item\label[claim]{G:1} If $u=v$, then $(u,v)_{s}=d_T(u,s).$
\item\label[claim]{G:2} $(\cdot,\cdot)_{s}$ is symmetric and non-negative.
\item\label[claim]{G:3} $(u,v)_{s}=0$ if and only if $s$ is on the unique path $[u,v]$.
\item\label[claim]{G:4} $(u,v)_{s} \leq d_T(u,s)$.
\end{enumerate}
\end{Lemma}
\begin{IEEEproof}
Both \cref{G:1} and \cref{G:2} follow immediately from definition (and the triangle inequality). For \cref{G:3}, $(u,v)_s=0$ if and only if $d_T(u,s)+d_T(v,s)=d_T(u,v)$. On a tree, this happens only when $s\in [u,v]$. \Cref{G:4} clearly holds if $(u,v)_s=0$. If $(u,v)_s > 0$, from \cref{G:3}, there exists a vertex $w \in [u,v]$ such that $[u,s]\cap [u,v] = [u,w]$ and $[v,s]\cap [u,v] = [v,w]$. Then,
\begin{dmath*}
	(u,v)_s = \ofrac{2}(d_T(u,w)+d_T(w,s) + d_T(v,w)+d_T(w,s) - d_T(u,w)-d_T(w,v))
	= d_T(w,s) \leq d_T(u,s),
\end{dmath*}
and the proof is complete.
\end{IEEEproof}
Let $U=\{u_1,\ldots,u_n\} \subset V$ be a subset of vertices of $G$, $s\notin U$ be a vertex of $G$, and $T$ be a tree that spans $U \cup \{s\}$ (i.e., $U \cup \{s\}$ is a subset of the vertex set of $T$). We define an $n\times n$ matrix $\bLambda$ whose $(i,j)$-th entry is the Gromov product $(u_i,u_j)_{s}$ in $T$. We call $\bLambda$ the \emph{Gromov matrix} with base $(T,s,U)$, where $T$ is known as its base tree and $s$ its base vertex. 

Recall that two trees $T_1$ and $T_2$ are said to be \emph{isometric} to each other if there is a bijection $\phi: T_1 \to T_2$ such that $d_{T_1}(u,v) = d_{T_2}(\phi(u),\phi(v))$ for any $u,v \in T_1$. The map $\phi$ is called an \emph{isometry}. We say that the two basis $(T_1,s_1,U_1)$ and $(T_2,s_2,U_2)$ are \emph{isometrically equivalent} to each other if there is an isometry $\phi: T_1 \to T_2$ such that $\phi(s_1)= s_2$ and $\phi(U_1)=U_2$. 

\begin{Proposition}\label{prop:unique}
Suppose that $\bLambda$ is a Gromov matrix, then its base $(T,s,U)$ is uniquely determined, up to isometric equivalence. 
\end{Proposition}
\begin{IEEEproof}
	We assume that $(T_1,s_1,U_1)$ and $(T_2,s_2,U_2)$ are two basis of $\bLambda$. We prove that they are isometrically equivalent to each other by induction on $|U_1|=|U_2|$. The statement is clearly true if $|U_1|=1$. Indeed, the condition that $T_i$ spans $U_i\cup \{s_i\}$, for $i=1,2$ imply that both $T_1$ and $T_2$ are simple paths, with end points $s_i$, and the single vertex in $U_i$, respectively. Therefore, $(T_1,s_1,U_1)$ and $(T_2,s_2,U_2)$ are equivalent as $T_1$ and $T_2$ have the same length given by $\bLambda$ (which is a $1\times 1$ matrix).
	
Suppose the statement holds for $|U_1|=|U_2|=n$. Consider the case where $|U_1|=|U_2|=n+1$. Let $\bLambda'$ be the $n\times n$ upper-left block of $\bLambda$. Since $\bLambda$ is a Gromov matrix, so is $\bLambda'$. For each $i=1,2$, there is a subset $U_i'\subset U_i$ with $|U_i'|=n$, and $T_i' \subset T_i$ such that $(T_i',s_i,U_i')$ is a base of $\bLambda'.$ By the induction hypothesis, $(T_i',s_i,U_i'), i=1,2$ are equivalent to each other via an isometry $\phi':T_1'\to T_2'.$ Without loss of generality, for $i=1,2$, let $U_i=\{u_{i,k} : k=1,\ldots,n+1\}$ such that $\phi'(u_{1,k})=u_{2,k}$, for $k=1,\ldots, n$.   
	
Let $1\leq j\leq n$ be an index such that the $(j,n+1)$-th entry $[\bLambda]_{j,n+1}$ of $\bLambda$ satisfy $[\bLambda]_{j,n+1}\geq [\bLambda]_{k,n+1}$ for all $1\leq k\leq n$. From \cref{lem:Gromov}\ref{G:4}, for each $i=1,2$, let $w_i$ be the unique vertex on the path from $u_{i,j}$ to $s_i$ such that $d_{T_i'}(w_{i},s_i)=[\bLambda]_{j,n+1}$. As $\phi'$ is an isometry and $\phi'(s_1)=s_2, \phi'(u_{1,j})=u_{2,j}$, we have $\phi'(w_1)=w_2.$
	
We claim that the path $[u_{i,n+1},s_i]$ in $T_i$ satisfies $[u_{i,n+1},s_1]\cap T_i' = [w_i,s_i]$. To see this, by the choice of $w_i$, we have 
\begin{align*}
[w_i,s_i]&=[u_{i,n+1},s_i]\cap [u_{i,j},s_i] \subset [u_{i,n+1},s_i]\cap T_i'.
\end{align*}	
On the other hand, the maximality on $[\bLambda]_{j,n+1}$ implies that $[u_{i,n+1},s_i]\cap T_i'$ cannot be a path longer than $[w_i,s_i]$. Hence, we have $[u_{i,n+1},s_i]\cap T_i' = [w_i,s_i]$. 
	
	Therefore, $(T_i\backslash T_i')\cup \{w_i\}=[w_i,u_{i,n+1}]$ for both $i=1,2$. Moreover, both $[w_1,u_{1,n+1}]$ and $[w_2,u_{2,n+1}]$ are of the same length $[\bLambda]_{n+1,n+1}-[\bLambda]_{j,n+1}$. Therefore, $\phi'$ can be extended to an isometry $\phi: T_1 \to T_2$ such that $\phi(u_{1,n+1})=u_{2,n+1}.$ The induction is now complete and the proposition is proved.
\end{IEEEproof}

For a base $(T,s,U)$, as the ordering of the vertices in the set $U$ is not fixed, its Gromov matrix $\bLambda$ is determined uniquely up to conjugation by permutation matrices of size $|U|$. With this observation and \cref{prop:unique}, we conclude that $\bLambda$ gives the same amount of information as $(T,s,U)$. Now we can describe a reformulation of the problem and the basic idea of our approach.

Since the propagation delays along each edge are continuous random variables, with probability one, the infection diffusion on a graph from a source $s$ forms a propagation path that is a spanning tree of all the nodes infected by $s$. Therefore, in the source inference task, in addition to the source $s$, we are implicitly required to find a $T$ in the set of spanning subtrees in $G$ of the observed infected set $\calV$ rooted at $s$. If $G$ is a tree, the source $s$ uniquely determines $T$. However, in a general graph, there may be more than one subtree rooted at $s$ that spans the observed infected nodes. Therefore, finding $(s, T)$ together is no longer equivalent to finding $s$ alone.  

In a dense graph, it is usually intractable to identify all possible $s$ and spanning subtrees rooted at $s$. On the other hand, the results in this section imply that to identify $s$ and the infection propagation tree $T$ is equivalent to estimating the Gromov matrix associated with the infection diffusion process. In this paper, we formulate the source estimation problem as an optimization of a cost function of a parametrized Gromov matrix and the diffusion parameters $(\mu,\sigma^2)$. By optimizing over a parametrized family of Gromov matrices, we overcome the intractability of identifying all possible Gromov matrices or infection propagation paths. However, the base tree of the optimal Gromov matrix found may not correspond to an actual subtree in $G$. Our proposed approach is therefore a heuristic in the case of general graphs. A widely used heuristic in the source inference literature \cite{Shah2010, zheng2015, Pinto2012, Louni2014} for general graphs is the BFS heuristic, which assumes that the infection propagation path is a BFS tree, which means that the infection is spread from a source to each observed infected node along a minimum-length path. However, this approximation has many limitations. For example, in general there exists more than one BFS tree rooted at a node, and the BFS heuristic typically chooses only one BFS tree \cite{Shah2010, zheng2015, Pinto2012, Louni2014}. Furthermore, the actual infection tree is not likely to be a BFS tree, especially when the graph is dense or $\sigma^2/\mu$ is large. In our Gromov matrix approach, some BFS trees are included in the family of Gromov matrices that we optimize over. Therefore, our method is expected to achieve better performance than the BFS heuristic.   

\section{Single source estimation}\label{sect:single source}

In this section, we consider the single source case, i.e., $S= \{s_1\}$. We first derive an MLE that simultaneously estimates the source and diffusion parameters when the network is a tree. We then describe our source estimation approach using Gromov matrices for general graphs. Finally, simulations on both synthetic and real-world networks are provided to compare the performance of our proposed algorithm with several other timestamp based source estimation algorithms in the literature.

\subsection{Preliminaries}
We first consider the special case where $G$ is a tree. Recall that $\calV=\{v_1,\ldots,v_n\}$ is a set of observed infected nodes with corresponding infection timestamps $\bT=[t_1,\ldots,t_n]'$. Suppose that $s_1$ is the source node. Let $t_0$ be the time the source node starts its infection spreading. Then, it is easy to see that the infection time of each node $v_k \in \calV$ is $t_k\sim \calN(t_0+d_G(s_1,v_k)\mu,d_G(s_1,v_k)\sigma^2)$, since all propagation delays $\{\tau_{uv}\}$ are i.i.d. $\calN(\mu,\sigma^2)$ variables. The covariance of $t_i$ and $t_j$ is $[\bLambda_{s_1}]_{i,j} \cdot \sigma^2$, with $\bLambda_{s_1}$ being the Gromov matrix with base $(T,s_1,\calV)$, where $T$ is the subtree rooted at $s_1$ that spans $\calV$. The likelihood of observing the infection times $\bT$ is then given by
	\begin{align}\label{eq:ml_pro}
	&p(\bT\mid s_1,t_0,\mu,\sigma^2)=\nonumber\\
	&\quad\quad{\frac{\exp\left(-\frac{1}{2\sigma^2}(\bT-\bD_{s_1}\bbeta_0)'\bLambda_{s_1}^{-1}(\bT-\bD_{s_1}\bbeta_0) \right)}{\sqrt{(2\pi\sigma^2)^n\det\bLambda_{s_1}}} },
	\end{align}
where $\bbeta_0=[t_0,\mu]'$, and $\bD_{s_1}$ is a matrix with its $k$-th row being $[1,d_G(s_1,v_k)]$. We have the following result.

\begin{Proposition}\label{Prop: sse_para}
Suppose that $G$ is a tree. Then, the MLE $(\hat{s},\hat{t}_0,\hat{\mu},\hat{\sigma}^2)=\argmax p(\bT\mid s_1,t_0,\mu,\sigma^2)$ for $(s_1,t_0,\mu,\sigma^2)$ is given by
\begin{align}\label{eq:sse_para}
\begin{aligned}
\hat{s} &= \argmin_{s\in V\backslash\calV} {\det(R_{s}\bLambda_{s}/n)},\\
[\hat{t}_0, \hat{\mu}]' &= \bbeta_{\hat{s}},\\
\hat{\sigma}^2 & =R_{\hat{s}}/n,
\end{aligned}
\end{align}
where 
\begin{align*}
R_s&=(\bT-\bD_s\bbeta_s)'\bLambda_s^{-1}(\bT-\bD_s\bbeta_s),\\
\bbeta_s&=(\bD_s'\bLambda_s^{-1} \bD_s)^{-1}\bD_s'\bLambda_s^{-1}\bT.
\end{align*}
\end{Proposition}
\begin{IEEEproof}
For any estimator $\hat{s}$ of $s_1$, by maximizing \cref{eq:ml_pro} w.r.t.\ $\bbeta_0=[t_0, \mu]'$, the MLE of $\bbeta_0$ is given by
	\begin{align*}
	\bbeta_{\hat{s}}
	&=\argmin_{\bbeta\in \Real^{2 \times 1}}(\bT-\bD_{\hat{s}}\bbeta)'\bLambda_{\hat{s}}^{-1}(\bT-\bD_{\hat{s}}\bbeta)\\
	&=(\bD_{\hat{s}}'\bLambda_{\hat{s}}^{-1} \bD_{\hat{s}})^{-1}\bD_{\hat{s}}'\bLambda_{\hat{s}}^{-1}\bT.
	\end{align*}
Similarly, the MLE of $\sigma^2$ is given by
	\begin{align*}
	\hat{\sigma}^2 &=\argmax_{\sigma^2\in\Real}{\frac{1}{\sqrt{(2\pi\sigma^2)^n}}\exp\left(-\frac{R_{\hat{s}}}{2\sigma^2}\right)}\\
	&=R_{\hat{s}}/n.
	\end{align*}
Finally, the MLE of $s_1$ is 
	\begin{align*}
	\hat{s}&=\argmax_{s\in V\backslash\calV}{p(\bT\mid s,\bbeta_s,\sigma^2=R_s/n)}\\
	&=\argmin_{s\in V\backslash\calV}{\det(R_s\bLambda_s/n)},
	\end{align*}	
	and the proof is complete.
\end{IEEEproof}

The reference \cite{Pinto2012} derives the MLE of the source $s_1$ by assuming that $(\mu,\sigma^2)$ are known, while \cref{Prop: sse_para} provides the MLE for both the source and diffusion parameters. 

\subsection{Single Source Estimation for General Graphs}\label{subsect:GSSI}

In this subsection, we consider source estimation for general graphs. To find the MLE for the source, we need to optimize the likelihood in \cref{eq:ml_pro} over all possible infection trees rooted at each candidate source node. As discussed in \cref{subsec:reformulation}, this is intractable due to multiple paths between each pair of vertices. Therefore, we propose to perform the optimization over a parameterized family of Gromov matrices as follows. We call the following procedure GSSI, which is summarized in \cref{Alg_GSSI}.
\begin{enumerate}[(i)]
	\item \label[step]{GSSI:1}
		For each candidate source node $s$, we first find two BFS trees. Let the two corresponding Gromov matrices be $\bLambda_s^1$ and $\bLambda_s^2$. For each $\theta\in[0,1]$, let
	\begin{align*}
	 \bM_s(\theta)&=\theta\bLambda_s^1+(1-\theta)\bLambda_s^2,\\    
	 \bbeta_s(\theta)&=(\bD_s'\bM_s^{-1}(\theta) \bD_s)^{-1}\bD_s'\bM_s^{-1}(\theta)\bT,
	\end{align*}
	and 
	\begin{align*}
	R_s(\theta)=(\bT-\bD_s\bbeta_s(\theta))'\bM_s^{-1}(\theta)(\bT-\bD_s\bbeta_s(\theta)).
	\end{align*}
	We find
	\begin{align}\label{eq:theta_s}
	\theta_s=\argmin_{\theta\in [0,1]}\det{(R_s(\theta)\bM_s(\theta)/n)},
	\end{align}
	and let $\bM_s\triangleq \bM_s(\theta_s)$.	
	\item \label[step]{GSSI:2}
	We introduce a target covariance matrix $\bH_s$, which is chosen to be  
	\begin{align}\label{eq:H}
		\bH_s\triangleq m_s\bI_n \text{ or } \bH_s\triangleq\diag(\bM_s),
	\end{align}
	where $m_s=\tr(\bM_s)/n$, $\bI_n$ is the identity matrix with dimension $n$, and $\diag(\bM_s)$ is the diagonal matrix with $[\diag(\bM_s)]_{i,i}=[\bM_s]_{i,i}$. This yields $\tr(\bH_s)=\tr(\bM_s)$. For each $\alpha\in[0,1]$, let
	\begin{align}
	\bA_s(\alpha)&=\alpha \bH_s+(1-\alpha)\bM_s,\label{eq:linear_comba}\\
	\tilde{\bbeta}_s(\alpha)&=(\bD_s'\bA_s^{-1}(\alpha) \bD_s)^{-1}\bD_s'\bA_s^{-1}(\alpha)\bT,
	\end{align}
	and
	\begin{align*} 
	\tilde{R}_s(\alpha)=(\bT-\bD_s\tilde{\bbeta}_s(\alpha))'\bA_s^{-1}(\alpha)(\bT-\bD_s\tilde{\bbeta}_s(\alpha)). 
	\end{align*}
  We find
	\begin{align}\label{eq:alpha_s}
	\alpha_s=\argmin_{\alpha\in [0,1]}\det{(R_s(\alpha)\bA_s(\alpha)/n)},
	\end{align}
  and let $\bA_s\triangleq \bA_s(\alpha_s)$.

	\item \label[step]{GSSI:3}
	We define our estimator $(\hat{s},\hat{t}_0,\hat{\mu},\hat{\sigma}^2)$ for $(s_1,t_0,\mu,\sigma^2)$ as
	\begin{align}\label{eq:sse_para_graph}
	\begin{aligned}
	\hat{s}&=\argmin_{s\in V\backslash \calV}{\det(R_s(\alpha_s)\bA_s/n)},\\
	[\hat{t}_0,\hat{\mu}]'&=\tilde{\bbeta}_{\hat{s}}(\alpha_{\hat{s}}),\\
	\hat{\sigma}^2 &= \tilde{R}_{\hat{s}}(\alpha_{\hat{s}})/n.
	\end{aligned}
	\end{align}	
\end{enumerate}

\Cref{GSSI:1,GSSI:2} above essentially construct a Gromov matrix to approximate the infection diffusion path starting at a node $s$, based on the observed infection timestamps $\bT$. In \cref{GSSI:2}, we let the new Gromov matrix $\bA_s$ be a convex combination of $\bM_s$ and a predefined invertible target matrix $\bH_s$ as shown in \cref{eq:linear_comba}. This construction is in the same spirit as the Stein-type shrinkage covariance matrix estimator proposed in \cite{Ledoit2004, touloumis2015}, which is used to overcome the ill-conditioning problem of the sample covariance matrix. If $\bM_s$ in \cref{eq:linear_comba} is replaced with the sample covariance matrix, then $\bA_s(\alpha)$  is the Stein-type covariance estimator, which can be written as a convex combination of the sample covariance matrix and a predefined invertible target matrix $\bH_s$. Two commonly used $\bH_s$ are shown in \cref{eq:H}. The parameter $\alpha$ is known as the shrinkage intensity, whose optimization can be found in works such as \cite{Ledoit2004, touloumis2015}. In \cref{GSSI:2}, we borrow the same idea; the difference is that we obtain the parameter $\alpha_s$ by solving the optimization problem in \cref{eq:alpha_s}. In this heuristic, the Gromov matrix $\bA_s$ obtained in \cref{GSSI:2} may not correspond to an actual subtree of $G$. However, in certain important special cases, for example, where either $\bH_s$ or $\bM_s$ is diagonal, then $\bA_s$ corresponds to a subtree of $G$. 

\begin{figure}
	\centering
	\includegraphics[scale=0.68]{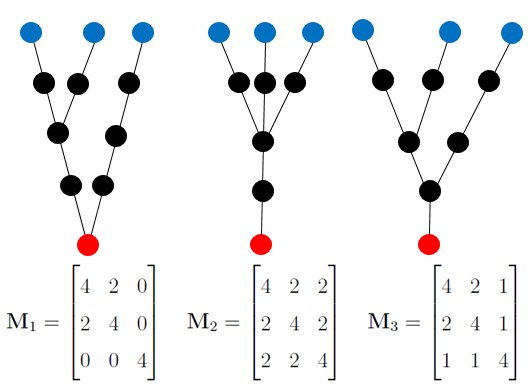}
	\caption{In this example, red nodes are the sources, blue nodes are the observed nodes. The Gromov matrices of the two trees on the left and middle are $\bM_1$ and $\bM_2$, respectively. The convex combination $\bM_3=(\bM_1+\bM_2)/2$ is the Gromov matrix of the tree on the right.}
	\label{fig:1}
\end{figure}

Forming convex combinations allows us to systematically approximate the Gromov matrices of spanning trees of $G$. In \cref{GSSI:1}, we propose to find the two BFS trees in opposite search directions: We first index the nodes from $1$ to $|V|$. Then we construct one BFS tree by prioritizing nodes with smaller indices, and another BFS tree by prioritizing nodes with bigger indices. Choosing BFS trees labeled in opposite directions may give us very distinct spanning trees; and hence we obtain a large family of Gromov matrices by taking convex combinations. We illustrate the geometric intuition by a simple example as in \cref{fig:1}. 

We now give an interpretation of \cref{GSSI:2}. 
When the graph is dense and the variance of the infection propagation delay is large, the spreading paths from the source to the observed infected nodes can be approximately regarded as uncorrelated (i.e., the spanning tree is a star tree). Therefore $\alpha$ in \cref{eq:linear_comba} indicates how far the spanning tree is from a star tree.
Intuitively, if the graph is denser, then $\alpha_{s_1}$ is greater, which is validated by the simulation results in \cref{fig:graph_ind} of \cref{subsect:exp_GSSI}. To demonstrate that \cref{GSSI:2} improves our estimation performance, we present simulations in \cref{subsect:exp_GSSI} where we compare the performance of GSSI with another procedure that we call naive-GSSI, which omits \cref{GSSI:2} and replaces $\bA_s$ in \cref{GSSI:3} with $\bM_s$. 

In \cref{GSSI:3}, the proposed estimator is simply the MLE assuming that $\bA_s$ is the Gromov matrix of the underlying infection tree. 

For each node in $V\backslash\calV$ we need to find two of their BFS trees, which can be computed with time complexity $O(|V|^2)$.\footnote{We say that $f(N) = O(g(N))$ if $f(N)/g(N)\leq k$ for some fixed $k>0$ as $N\to\infty$.} The optimization problems in \cref{eq:theta_s} and \cref{eq:alpha_s} do not have analytical solutions. We utilize standard iterative techniques like gradient descent to solve them, which require a time complexity of $O(|\calV|^3)$, assuming that the maximum number of iterations used is fixed. The overall time complexity of GSSI is then $O(|V|(|V|^2 + |\calV|^3)) = O(|V|^3 + |V||\calV|^3)$. To compare with other algorithms, the time complexity of TRBS  proposed in \cite{Shen2016} is $O(|\calV||V|^2)$; and the time complexity of the method proposed in \cite{Pinto2012} is $O(|V|^3+ |V||\calV|^3)$ (note that $|V|$ matrix inverses are required in \cite{Pinto2012}, each incurring a time complexity of $O(|\calV|^3)$), assuming that the diffusion parameters are known a priori.

\begin{algorithm}[!htb]
	\caption{Gromov Single Source Identification (GSSI)}
	\label{Alg_GSSI}
	\begin{algorithmic}[1]		
		\REQUIRE Adjacency matrix of the graph $G$, $\calV=\{v_1,\ldots,v_n\}$ and $\bT=[t_1,\ldots,t_n]'$. 
		\ENSURE $(\hat{s},\hat{t}_0,\hat{\mu},\hat{\sigma}^2)$
		\STATE Let $f_m=\infty$.
		\FOR {every $s\in V\backslash\calV$}
		\STATE Find two BFS trees rooted at $s$ in opposite search directions to obtain $\bLambda_s^1$ and $\bLambda_s^2$.
		\STATE Find $\theta_s$ according to \cref{eq:theta_s} to obtain $\bM_s$.
		\STATE Find $\alpha_s$ according to \cref{eq:alpha_s} to obtain $\bA_s$.
		\STATE\label[line]{GSSI:fs} Compute $f_s=\det(R_s(\alpha_s)\bA_s/n)$.
		\IF {$f_s\leq f_m$}
		\STATE Set $f_m=f_s$, and $\hat{s}=s$.
		\ENDIF
		\ENDFOR
		\STATE Set $(\hat{t}_0,\hat{\mu},\hat{\sigma}^2)$ according to \cref{eq:sse_para_graph}.
	\end{algorithmic}
\end{algorithm}

\subsection{Discussion}\label{subsect:unity}

In the GSSI algorithm, if $G$ is a tree then we do not need to optimize $\theta$ because $\bLambda_s^1=\bLambda_s^2$, but in this case, GSSI is not equivalent to MLE due to the introduction of the parameter $\alpha_s$ and target matrix $\bH_s$. However, simulations in \cref{subsect:exp_GSSI} indicate that the introduction of $\alpha_s$ and $\bH_s$ does not unreasonably impair our estimation results in trees, while the introduction of the additional step in \cref{eq:alpha_s} leads to better performance for general graphs. 

Simulations in \cref{subsect:exp_GSSI} suggest that when $G$ is a tree, then with high probability the MLE in \cref{Prop: sse_para} and GSSI give the same source estimates, and $\alpha_{\hat{s}}$ obtained by GSSI is close to 0, which implies that the performance of GSSI is close to the MLE when the graph is a tree. We provide some theoretical results in \cref{app:heuristic} that indicate why this is true.

\subsection{Simulations for Single Source Estimation}\label{subsect:exp_GSSI} 

In this subsection, we present simulation results on both synthetic and real networks to compare the performance of GSSI with the MLE in \cref{Prop: sse_para}, the TRBS algorithm proposed in \cite{Shen2016}, and the method proposed in \cite{Pinto2012}, which we call the GAU algorithm. We note that both the TRBS and GAU algorithms require prior knowledge of the mean and/or variance of the propagation delay along each graph edge, while our approach estimates these parameters from the observed timestamps. 

We first perform simulations on two kinds of random trees. Starting from one node, we add a new node in every step and attach it to one of the existing nodes with probabilities proportional to their degrees to obtain a scale-free tree, or attach the new node to one of the existing nodes randomly to obtain a non-scale-free tree. We call these the Barab\'{a}si-Albert (B-A) tree and Erd\H{o}s-R\'{e}nyi (E-R) tree, respectively. We denote B-A trees as $\text{BA}(N)$ and E-R trees as $\text{ER}(N)$, where $N$ is the number of nodes. 
Some properties like the diameter and average pairwise distance are listed in Table \ref{tab:graph_properties}. 
Our simulations indicate that choosing $\bH_s=m_s\bI_n\text{ or } \diag(\bM_s)$ in \cref{eq:H} does not lead to much difference in the performance of GSSI. Therefore, here we only present the results for $\bH_s=m_s\bI_n$.
\begin{table}[!htbp]
	\centering  
	\begin{tabular}{lccc}  
		\hline
		Graph &$|E|/|V|$ &Diameter &\makecell*[l]{Average pairwise \\ distance}\\ 
		\hline  
		$\ER(500)$ &1.00 &20 &9.23  \\        
		$\BA(500)$ &1.00 &15 &6.40 \\   
		$\ER(500,4)$ &1.98 &11 &4.69 \\  
		$\BA(500,4)$ &1.99 &7 &3.82 \\  
		$\ER(500,16)$ &8.12 &4 &2.54 \\  
		$\BA(500,16)$ &7.88 &4 &2.46 \\   
		Enron &9.86 &9 &3.32\\    
		Facebook &25.9 &9 &2.95 \\ 
		Twitter &5.21 &10 &3.41 \\
		\hline
	\end{tabular}
	\caption{Some graph properties of the networks used in our simulations.}
	\label{tab:graph_properties}
\end{table}

For each simulation, we randomly pick a node from the network to be the source. We simulate the propagation delays along each edge using a truncated Gaussian distribution $\calT\calN (2,1)$, and randomly choose a subset of nodes as observed nodes. We perform 300 simulation runs. The error distance is defined as the distance between the estimated source and the real source. All four algorithms rank all the nodes in the graph according to their likelihood of being the source. Following \cite{Zhu2014}, we define $\gamma$\%-accuracy as the proportion of simulations in which the real sources are ranked in the top $\gamma$ percent of all the nodes. 

\begin{figure}[!htb]
	\centering
	\begin{minipage}[b]{.5\linewidth}
		\centering
		\centerline{\includegraphics[width=4.3cm]{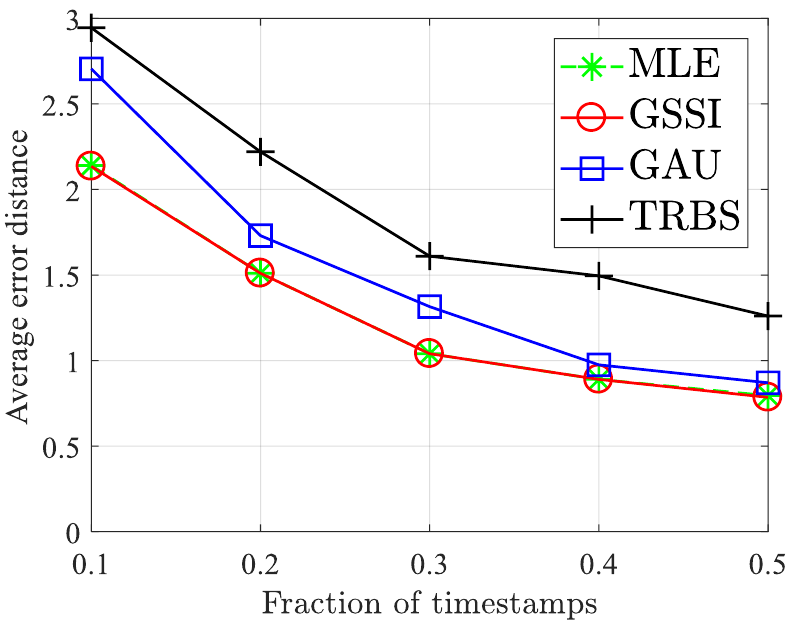}}
		\centerline{(a) $\ER(500)$}
	\end{minipage}%
	\begin{minipage}[b]{.5\linewidth}
		\centering
		\centerline{\includegraphics[width=4.3cm]{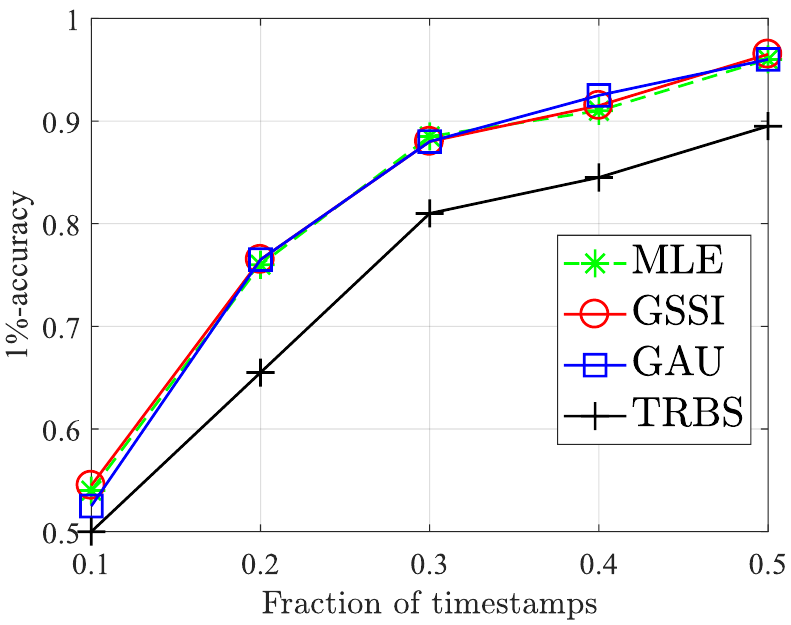}}
		\centerline{(b) $\ER(500)$}
	\end{minipage}
	\begin{minipage}[b]{.5\linewidth}
		\centering
		\centerline{\includegraphics[width=4.3cm]{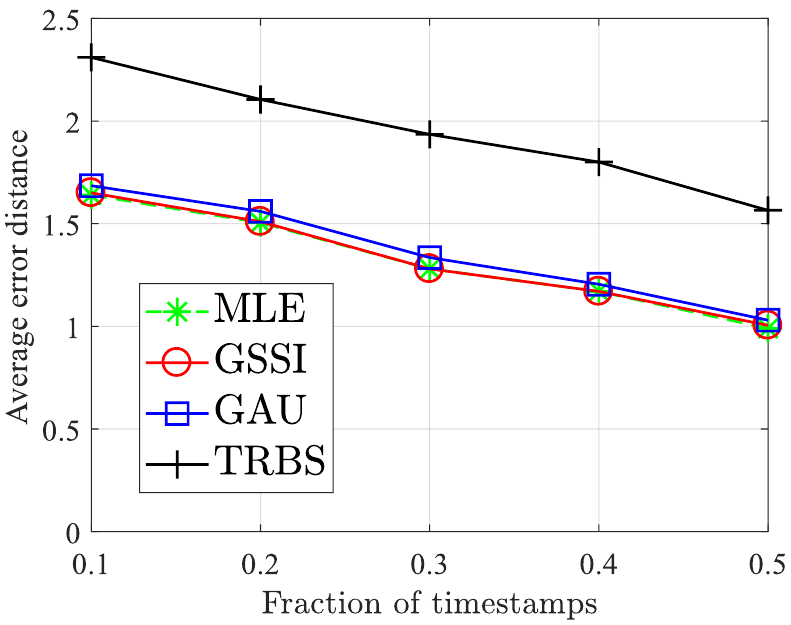}}
		\centerline{(c) $\BA(500)$}
	\end{minipage}%
	\begin{minipage}[b]{.5\linewidth}
		\centering
		\centerline{\includegraphics[width=4.3cm]{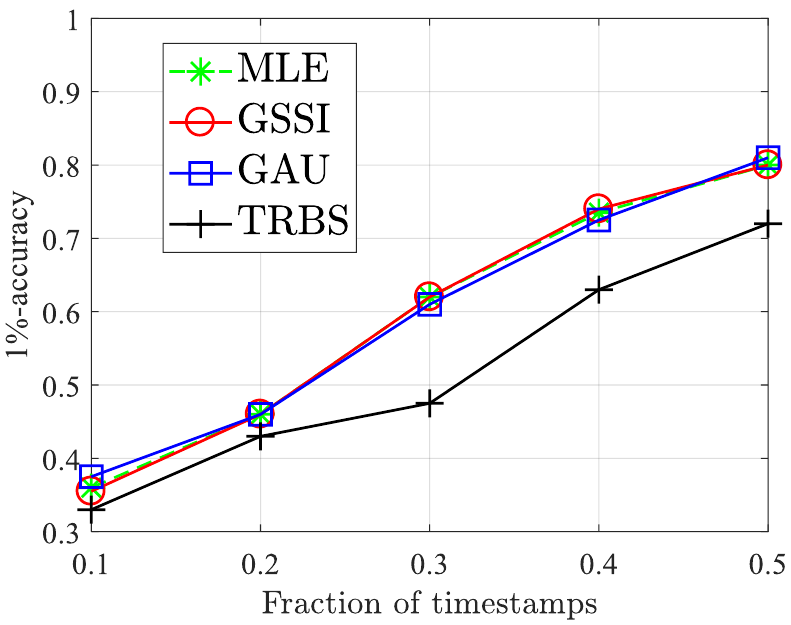}}
		\centerline{(d) $\BA(500)$}
	\end{minipage}
	\caption{Single source estimation on $\ER(500)$ and $\BA(500)$. }
	\label{fig:tree_compare}
\end{figure}

\begin{figure}[!htb]
	\centering 
	\begin{minipage}[b]{0.5\linewidth}
		\centering
		\includegraphics[width=4.3cm]{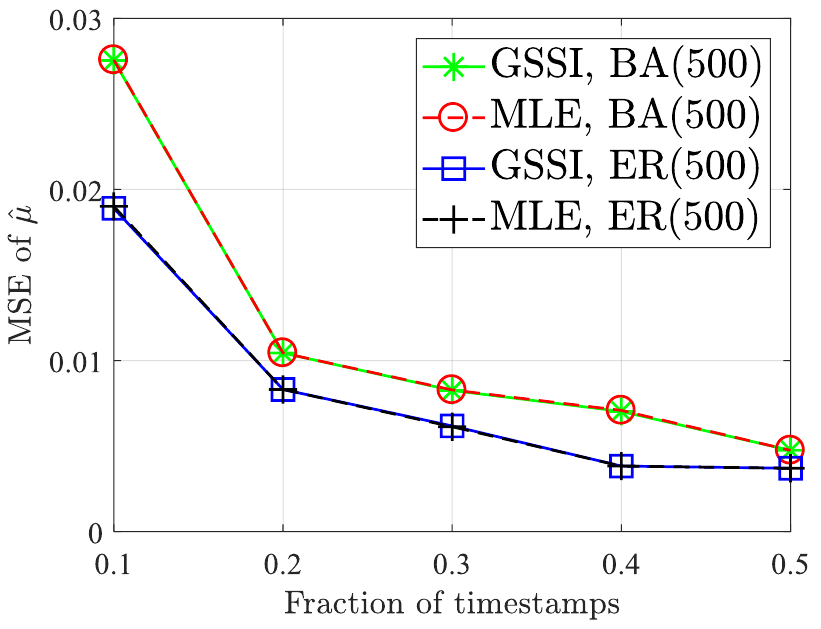}
	\end{minipage}%
	\begin{minipage}[b]{0.5\linewidth}
		\centering
		\includegraphics[width=4.3cm]{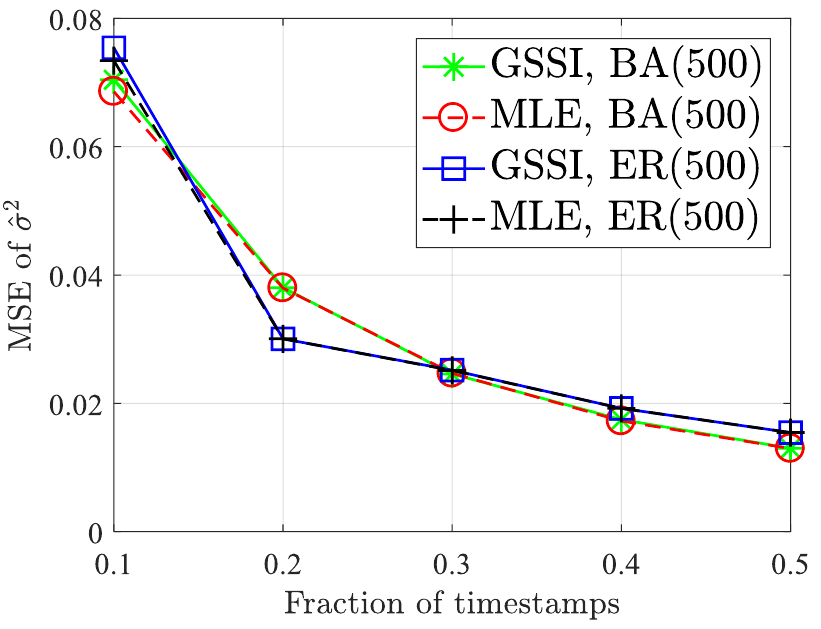}
	\end{minipage}
	\caption{MSE of $\hat{\mu}$ and $\hat{\sigma}^2$ obtained by MLE and GSSI.}
	\label{fig:MSE}
\end{figure}

Simulation results are shown in \figref{fig:tree_compare}. We only consider the case where the fraction of timestamps is less than 50\%. Simulations indicate that the performance of GSSI is very close to the MLE when $G$ is a tree, which means that the introduction of parameter $\alpha_s$ and target matrix $\bH_s$ in \cref{eq:linear_comba} does not unreasonably impair our estimation results when $G$ is a tree. 
We observe that GSSI performs no worse than GAU in almost all cases, and better than TRBS for both kinds of random trees although GSSI requires less information. The reason why GAU performs worse than GSSI in most cases is because the simulated propagation delays are generated using a truncated Gaussian distribution instead of the Gaussian distribution assumed by GAU. Although GSSI also assumes a Gaussian distribution, it estimates the distribution parameters from the observed data, which help to mitigate this mismatch. TRBS has the worst performance because it does not utilize any variance information. For GSSI, we compute the mean squared error (MSE) of the estimates $\hat{\mu}$ and $\hat{\sigma}^2$ w.r.t.\ $\mu=2$ and $\sigma^2=1$ respectively in \cref{fig:MSE}, which show that the MSE of GSSI is very close to that of the MLE. 

\begin{figure}[htbp]
	\footnotesize
	\centering
	\begin{minipage}[b]{.5\linewidth}
		\centering
		\centerline{\includegraphics[width=4.3cm]{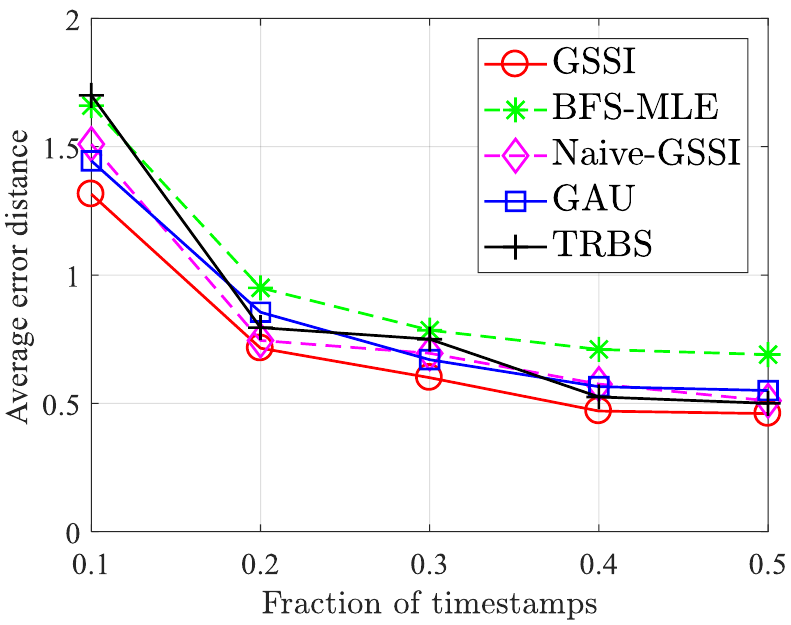}}
		\centerline{(a) $\ER(500,4)$}
	\end{minipage}%
	\begin{minipage}[b]{.5\linewidth}
		\centering
		\centerline{\includegraphics[width=4.3cm]{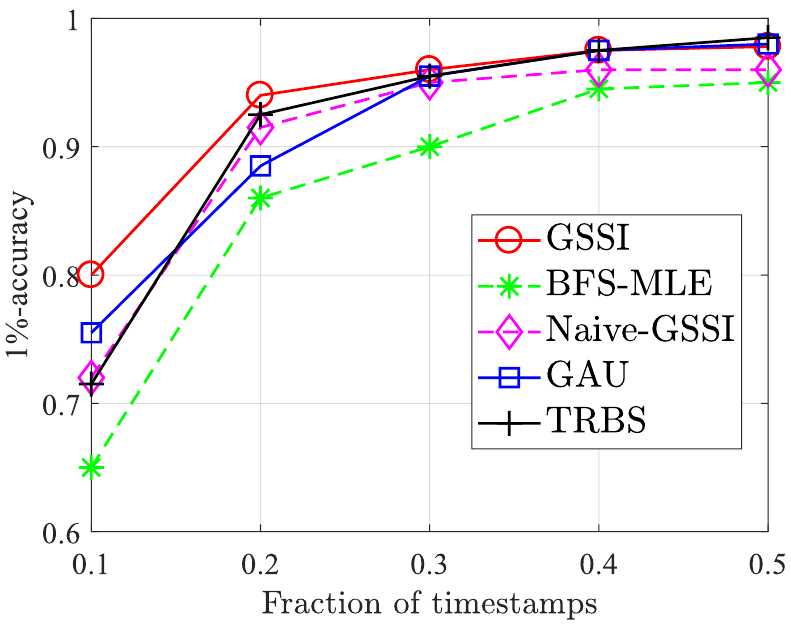}}
		\centerline{(b) $\ER(500,4)$}
	\end{minipage}
	\begin{minipage}[b]{.5\linewidth}
		\centering
		\centerline{\includegraphics[width=4.3cm]{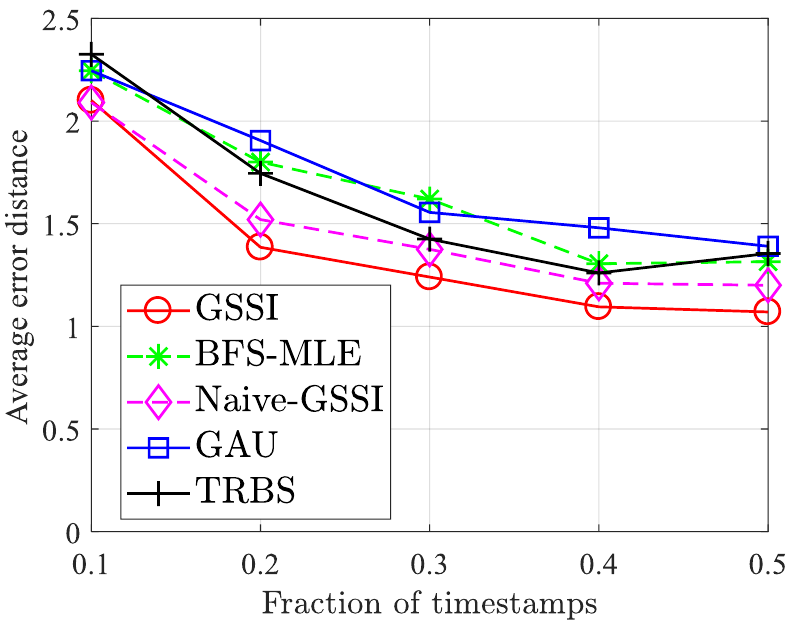}}
		\centerline{(c) $\BA(500,4)$}
	\end{minipage}%
	\begin{minipage}[b]{.5\linewidth}
		\centering
		\centerline{\includegraphics[width=4.3cm]{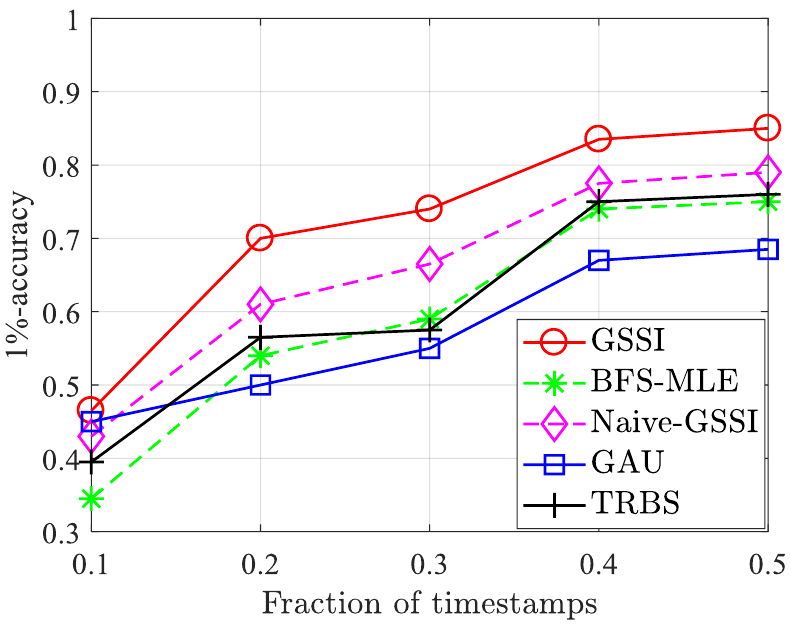}}
		\centerline{(d) $\BA(500,4)$}
	\end{minipage}
	\caption{Single source estimation on $\ER(500,4)$ and $\BA(500,4)$. For the diffusion process, set $(\mu,\sigma^2)=(2,1)$.}
	\label{fig:graph_4_compare}
\end{figure}

\begin{figure}[htbp]
	\footnotesize
	\centering
	\begin{minipage}[b]{.5\linewidth}
		\centering
		\centerline{\includegraphics[width=4.3cm]{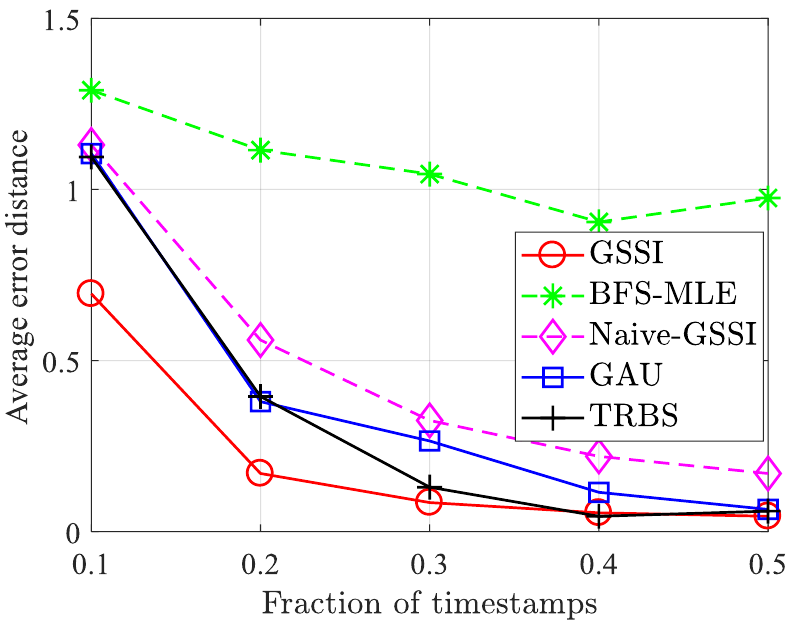}}
		\centerline{(a) $\ER(500,16)$}
	\end{minipage}%
	\begin{minipage}[b]{.5\linewidth}
		\centering
		\centerline{\includegraphics[width=4.3cm]{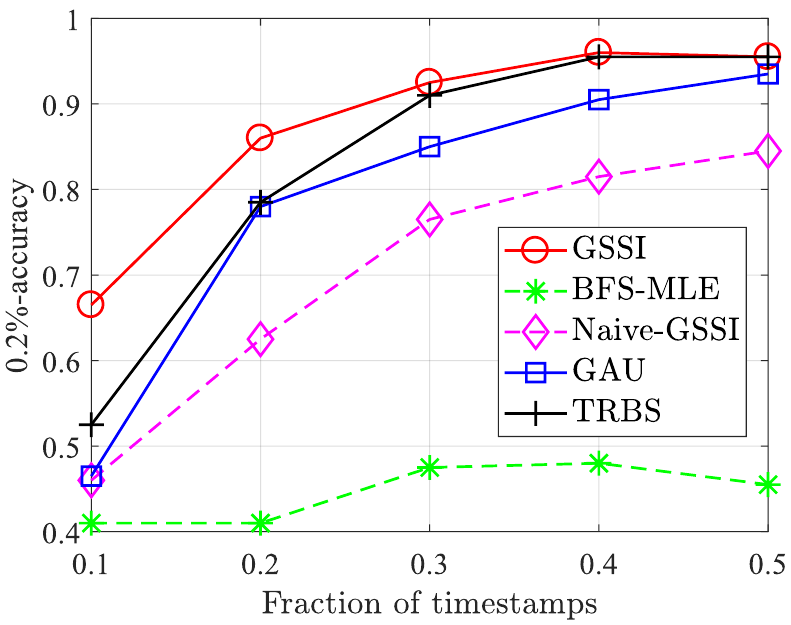}}
		\centerline{(b) $\ER(500,16)$}
	\end{minipage}
	\begin{minipage}[b]{.5\linewidth}
		\centering
		\centerline{\includegraphics[width=4.3cm]{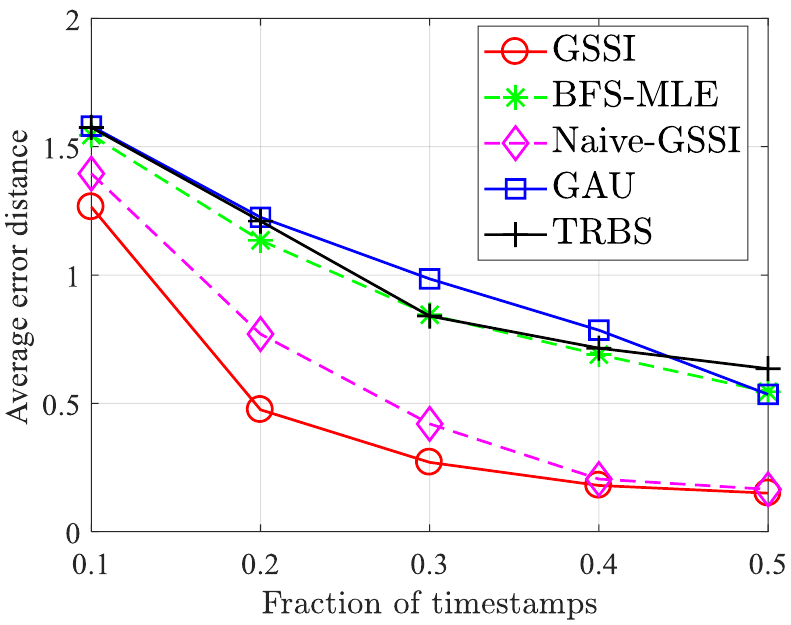}}
		\centerline{(c) $\BA(500,16)$}
	\end{minipage}%
	\begin{minipage}[b]{.5\linewidth}
		\centering
		\centerline{\includegraphics[width=4.3cm]{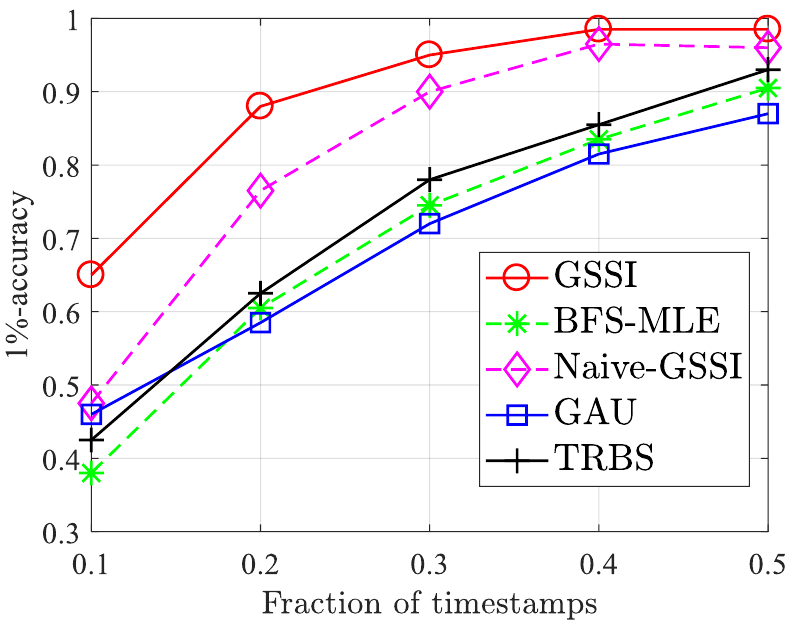}}
		\centerline{(d) $\BA(500,16)$}
	\end{minipage}
	\caption{Single source estimation on $\ER(500,16)$ and $\BA(500,16)$, with $(\mu,\sigma^2)=(3,1)$.}
	\label{fig:ER_16_compare}
\end{figure}

\begin{figure}[htbp]
	\footnotesize
	\centering
	\begin{minipage}[b]{.5\linewidth}
		\centering
		\centerline{\includegraphics[width=4.3cm]{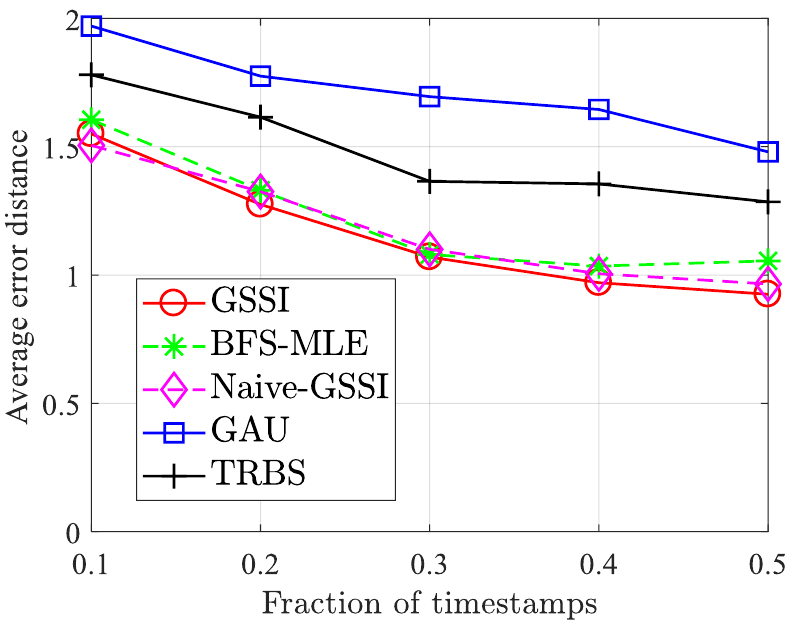}}
		\centerline{(a) Enron network}
	\end{minipage}%
	\begin{minipage}[b]{.5\linewidth}
		\centering
		\centerline{\includegraphics[width=4.3cm]{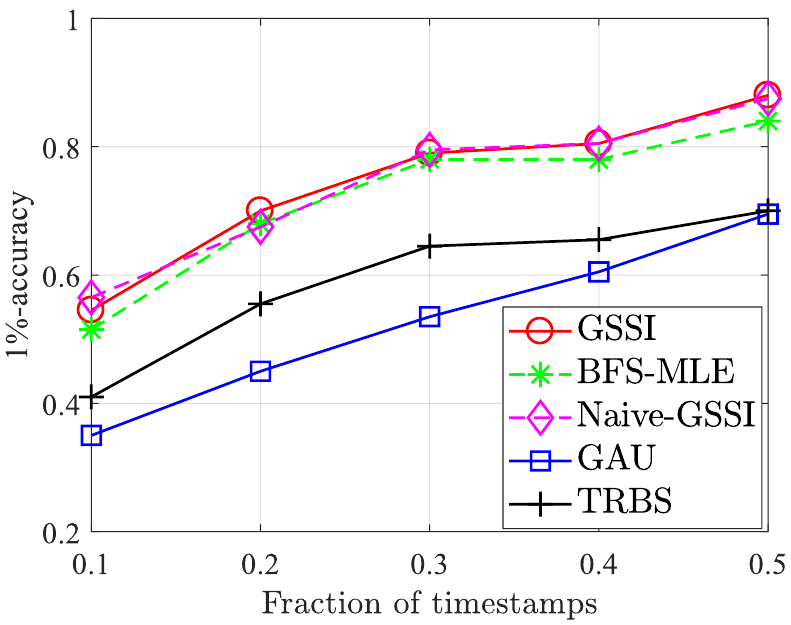}}
		\centerline{(b) Enron network}
	\end{minipage}
	\begin{minipage}[b]{.5\linewidth}
		\centering
		\centerline{\includegraphics[width=4.3cm]{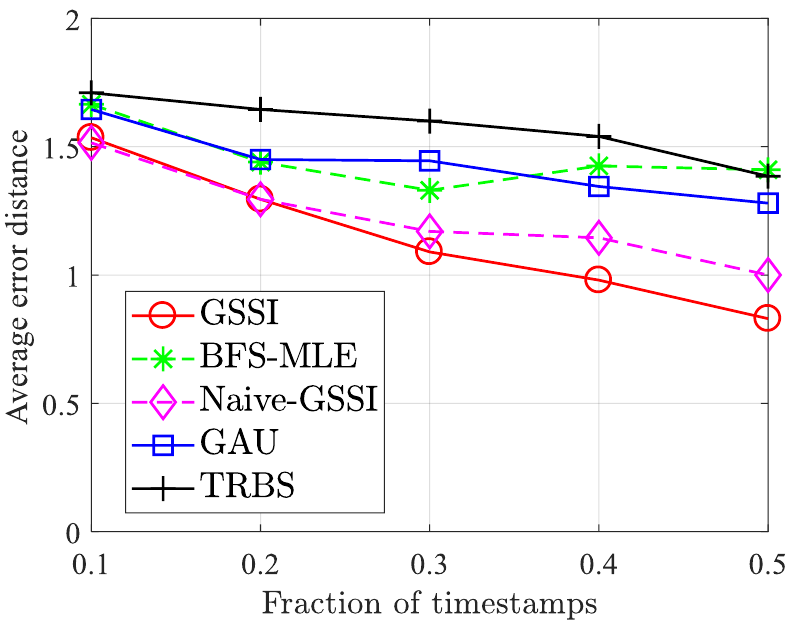}}
		\centerline{(c) Facebook network}
	\end{minipage}%
	\begin{minipage}[b]{.5\linewidth}
		\centering
		\centerline{\includegraphics[width=4.3cm]{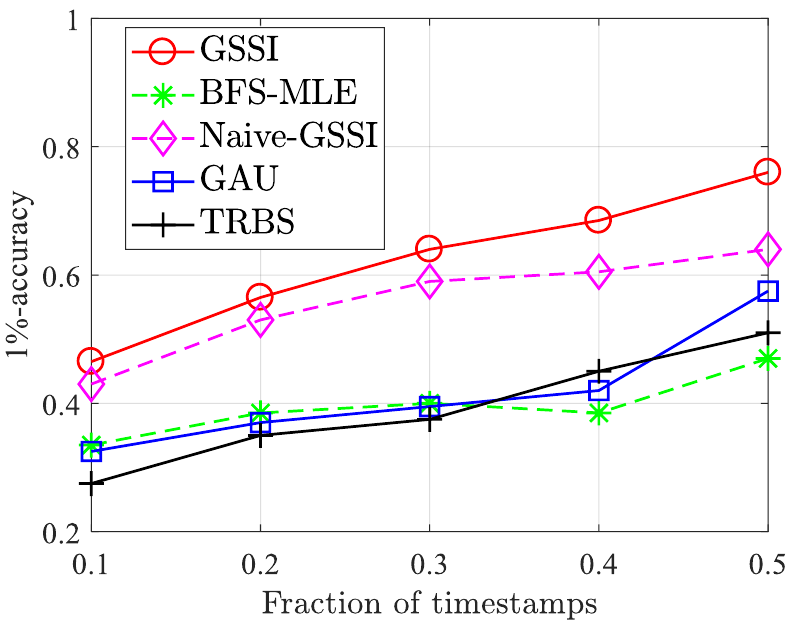}}
		\centerline{(d) Facebook network}
	\end{minipage}
	\caption{Single source estimation on two real networks, with $(\mu,\sigma^2)=(3,1)$.}
	\label{fig:real_network_compare}
\end{figure}

We next perform simulations on general B-A graphs $\BA(N,Np)$ proposed in \cite{BA_random} and E-R graphs $\ER(N,Np)$ proposed in \cite{ER_random}, where $N$ is the number of nodes, and $Np$ is the expected degree of each node. B-A graphs are scale-free while E-R graphs are not. We choose $N=500$, and $Np=4$ and $16$ to simulate graphs with different densities. We also test on two real networks including the Enron email network\footnote{https://snap.stanford.edu/data/email-Enron.html} and a Facebook network\footnote{https://snap.stanford.edu/data/egonets-Facebook.html} provided by SNAP. For the Enron email network, we extract a subgraph with 670 nodes and 3303 edges, and for the Facebook network, we extract a subgraph with 1034 nodes and 26749 edges. See Table \ref{tab:graph_properties} for some graph properties. To demonstrate that the introduction of the parameters $\theta_s, \alpha_s$ and $\bH_s$ can improve performance, we also compare GSSI with two additional algorithms which we call BFS-MLE and naive-GSSI. For BFS-MLE we simply use the BFS-tree heuristic: For each node $s\in V\backslash\calV$, we find a BFS tree rooted at $s$ and then perform MLE to compute $g_s\triangleq\det(R_s\bLambda_s/n)$ from \cref{Prop: sse_para}. Finally, we minimize $g_s$ over $s\in V\backslash\calV$ to obtain the source estimate. For the naive-GSSI, we follow the same procedure as GSSI, but only optimize $\theta_s$ in \cref{eq:theta_s}, and let $\alpha_s=0$.  Simulation results are shown in \cref{fig:graph_4_compare,fig:ER_16_compare,fig:real_network_compare}. The plots show that in almost all the cases,  GSSI performs best and achieves significant improvement over BFS-MLE, while naive-GSSI also performs better than BFS-MLE. Therefore, we see that the introduction of $\theta_s, \alpha_s$ and $\bH_s$ in GSSI can improve estimation performance.

\begin{figure}[!htb]
	\centering
		\centering
		\centerline{\includegraphics[width=5.5cm]{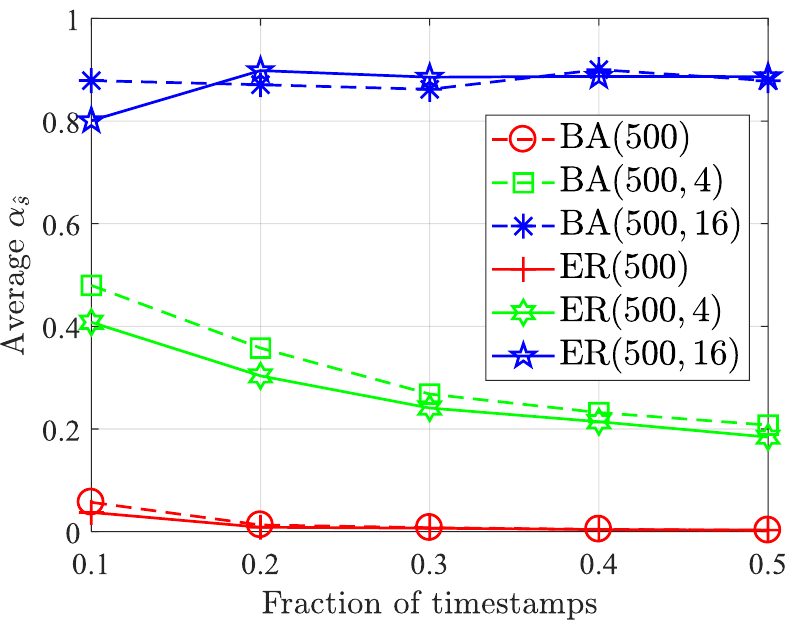}}
	\caption{Simulations to study how $\alpha_{\hat{s}}$ varies with the fraction of
		observed timestamps and for different graphs.}
	\label{fig:graph_ind}
\end{figure}

We also study how $\alpha_{\hat{s}}$ in \cref{eq:alpha_s} varies with increasing fraction of observed timestamps. From \cref{fig:graph_ind}, we see that for two kinds of random trees, the average $\alpha_{\hat{s}}$ is very close to 0, which agrees with the intuition provided by \cref{lem:alpha_opt} in \cref{app:heuristic}. For B-A and E-R graphs, the average $\alpha_{\hat{s}}$ increases as the average degree of the graph gets larger, which is intuitively satisfying because the actual infection propagation path can deviate significantly from BFS trees as the graphs become denser. 

\begin{figure}[!htb]
	\centering 
	\begin{minipage}[b]{0.5\linewidth}
		\centering
		\includegraphics[height=3.45cm]{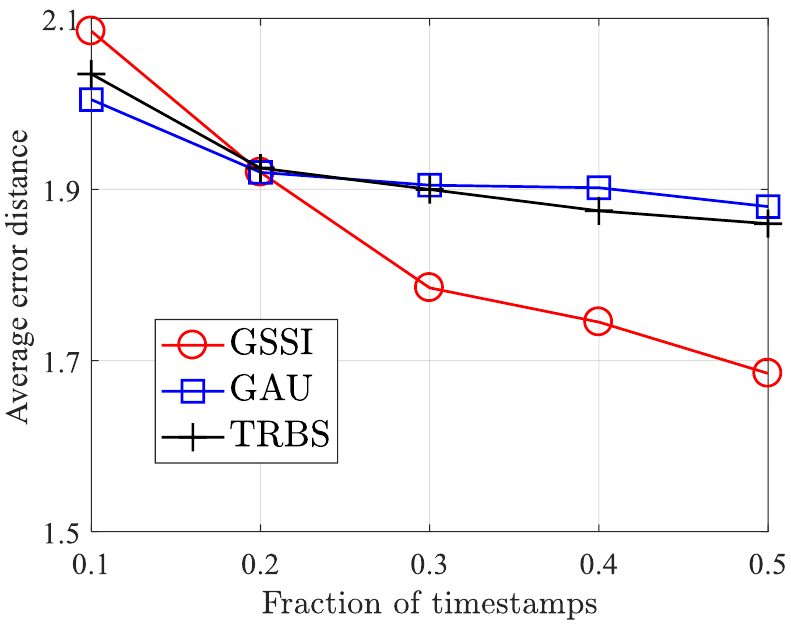}
	\end{minipage}%
	\begin{minipage}[b]{0.5\linewidth}
		\centering
		\includegraphics[height=3.45cm]{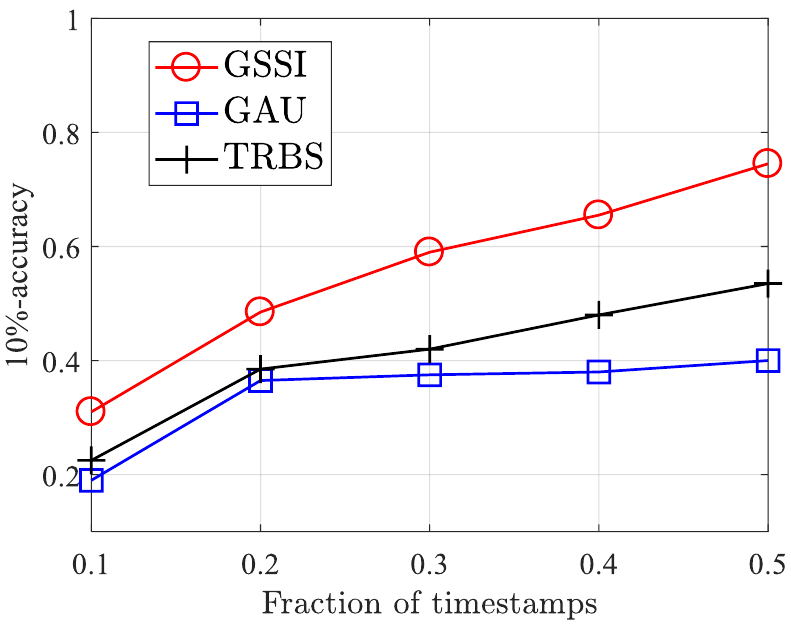}
	\end{minipage}
	\caption{Malware source estimation on Facebook network.}
	\label{fig:malware_GSSI}
\end{figure}

Finally, we apply our GSSI algorithm to find the source of malware propagation in online social networks. Malware attacks have become ubiquitous in online social networks such as Facebook and LinkedIn \cite{FB2015, FB2016}. A successful attack using malware in a network can result in tens of millions of accounts being compromised and users' computers being infected. For example, Trojan, as the most popular type of malware, is able to steal confidential information, install ransomware and infect other computers in the network. Therefore, it is important to identify the sources of Trojans in order to analyse and remove them as early as possible. There have been many works studying the propagation of Trojans\cite{FAN2011,Yan2011,Faghani2012,Faghani2017}. In this paper, we adopt the propagation model in \cite{Faghani2017}, which is a spatial-temporal SIR model that takes into account the network topology and temporal dynamics of user activities. Furthermore, it considers characteristics of modern Trojans, security practices, and user behaviors. Similar to the simulations in \cite{Faghani2017}, we use the Facebook network subgraph described above. Parameters used for the propagation model are shown in \cref{tab:malware} (see \cite{Faghani2017} for more details). We assume that the malware is a zero-day Trojan, which propagates fast. Therefore, it has a high $p_i$, and small $q_i$ and $\delta_i$. The parameter $\beta_{\text{max}}$ is set according to a survey conducted by Microsoft \cite{survey2015}. Since GAU and TRBS require knowledge of the mean propagation delays, we average the infection propagation times over the edges to obtain an estimate. Simulation results are shown in \cref{fig:malware_GSSI}. The plots show that although the propagation model violates our Gaussian assumption, GSSI still performs best in most cases and is able to estimate the true malware source within 1.92 hops with at least 20\% infection timestamps.

\begin{table}[!htb]
	\centering  
	\begin{tabular}{|l|p{0.68\columnwidth}|c|}  
		\hline
		Parameter &Description\\
		\hline 
		$p_i=0.8$ &Probability of user $i$ following a malicious post, unknowingly downloading the malware and executing it.\\
		\hline
		$\delta_i=0.2$ &Probability of user $i$ accepting clean-up solutions from his non-infectious friends.\\
		\hline
		$q_i=0.2$ &Probability of user $i$ recovering independently without assistance from his friends.\\
		\hline
		$\beta_{\text{max}}=0.75$ &Percentage of the social network population that has anti-virus products installed on their computers.\\
		\hline
		$\tau_i\sim\Exp(0.2)$ &User message checking time.  \\
		\hline
	\end{tabular}\\
	\caption{Parameter description and setting for malware propagation model of \cite{Faghani2017}.}
	\label{tab:malware}
\end{table}

To summarize, we see that the estimation accuracy for all algorithms is influenced by the network topology, diffusion model and fraction of timestamps. Generally it is costly to observe too many nodes, so we may need to estimate the source to within a small number of hops of the real source with a small fraction of observed timestamps. In all our experiments, GSSI is able to estimate the source on average to within 1.8 hops with only 30\% of timestamps. A summary comparison of GSSI w.r.t.\ the next best algorithm used in our experiments when 30\% timestamps are observed is provided in \cref{tab:GSSI_compare}.  

\begin{table}[!htbp]
	\centering  
	\begin{tabular}{|l|c|c|}  
		\hline
		Graph & Error reduction(\%)\\
		\hline   
		$\ER(500)$ & 21.2 \\       
		$\BA(500)$ & 4.5 \\ 
		$\ER(500,4)$ &10.4  \\ 
		$\BA(500,4)$ &13.0  \\   
		$\ER(500,16)$ &38.5 \\ 
		$\BA(500,16)$ &67.9 \\  
		Enron & 21.6\\   
		Facebook & 24.8 \\ 
		Facebook (malware)& 6.1\\
		\hline
	\end{tabular}
	\caption{Percent of reduced average error distance GSSI compared with the next best algorithm.}
	\label{tab:GSSI_compare}
\end{table}

\section{Multiple sources estimation}\label{sect:multiple sources}

In this section, we discuss the case of multiple infection sources, i.e., $|S|>1$ but unknown. We first reduce the multiple sources estimation problem to the single source estimation problem, which is then solved by applying GSSI. We develop a reduction step for trees, and then extend it heuristically to general graphs.

\subsection{The Reduction Step for Trees}\label{sect:Reduction}

\begin{figure}[!htb]
	\centering 
	\subfigure[]{
		\begin{minipage}[b]{1\linewidth}
			\centering
			\includegraphics[width=4.5cm]{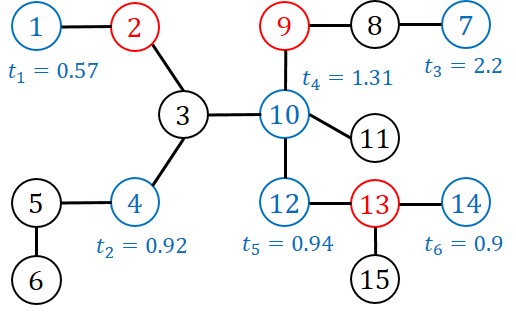}
		\end{minipage}
		\label{fig:obs}
	}
	\subfigure[]{
		\begin{minipage}[b]{1\linewidth}
			\centering
			\includegraphics[width=6.3cm]{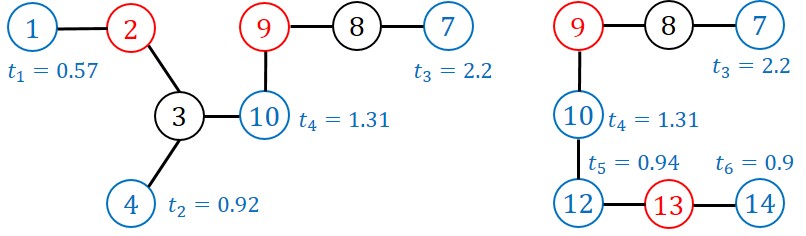}
		\end{minipage}
		\label{fig:obs_MSR}
	}
	\subfigure[]{
		\begin{minipage}[b]{1\linewidth}
			\centering
			\includegraphics[width=6.3cm]{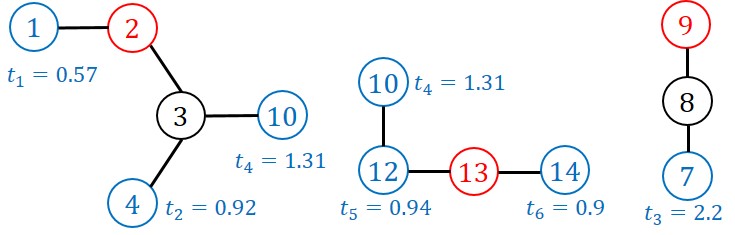}
		\end{minipage}
		\label{fig:MSR_SSSE}
	}
	\captionsetup{singlelinecheck=off}
	\caption{An example to illustrate the notions of observation cluster, source candidate cluster and the SCCE algorithm. 
	\begin{enumerate}[(a)]
		\item Red nodes are the sources, blue nodes are the observed nodes. Three infection sources $S=\{2,9,13\}$ initiate a diffusion process on the tree shown. Suppose we observe 6 infected nodes $\calV=\{1,4,7,10,12,14\}$ and their respective timestamps. For node 9, since $t_{4}$ is greater than $t_{1}$, $t_{2}$ and $t_{5}$, the observation cluster for node 9 is $\calV_9=\{7,10\}$. Similarly, $\calV_8=\calV_{10}=\{7,10\}$, and for any node $u\notin\{8,9,10\}$, we have $\calV_u\neq \{7,10\}$. Therefore, the source candidate cluster w.r.t. $\calV_9$ is $A_9=\{8,9,10\}$. 
		\item In the MSR step of SCCE, we first set $\calU=\calV$, and obtain $\xi_1=\argmin_{v_k\in\calU} t_k = 1$ so that $\calV_{\xi_1}=\calV_1=\{1,4,7,10\}$ and $A_{\xi_1}=A_1=\{1,2,3\}$. We next set $\calU=\calU\backslash\calV_{\xi_1}=\{12,14\}$, and obtain $\xi_2=\argmin_{v_k\in\calU} t_k =14$ with $\calV_{\xi_2}=\calV_{14}=\{7,10,12,14\}$ and $A_{\xi_2}=A_{14}=\{13,14\}$. We then have $l=2$ observation clusters, and the subtrees spanning $\calV_{\xi_1}$ and $\calV_{\xi_2}$ are shown in (b). Note that $A_{\xi_1}$ and $A_{\xi_2}$ each contains at least one source, and $l < |S|=3$ since $\nu_9=10\in\calV_{13}$. This is consistent with \cref{Prop: covering}.
		\item In the SSSE step of SCCE, for each subtree in (b), we first find a maximum length path $P$, and define $s_P(w)$ to be the sum of the distances between $w$ and leaves that are closest to $w$ compared with other nodes in $P\backslash w$ . For the left subtree in (b), we find $P=[1,7]$ with the maximal length 6. We have $s_P(w)=1$ when $w\in\{2,3,8\}$, and $s_P(w)=0$ when $w\in\{9,10\}$. Since $s_P(9)$ is a local minimum, we delete the edge connecting nodes 9 and 10 to obtain two subtrees as shown in the left and middle of (c). For the right tree in (b), we have $P=[7,14]$, $s_P(w)=1$ when $w\in\{8,13\}$, and $s_P(w)=0$ when $w\in\{9,10,12\}$. We delete the edge connecting nodes 9 and 10 and obtain two subtrees as shown in the middle and right of (c). 
	\end{enumerate}}\label{fig:obs_adm} 
\end{figure}

In this subsection, we consider the case where $G$ is a tree, and introduce the concepts of observation cluster and source candidate cluster, which allow us to reduce the multiple sources estimation problem to that of estimating a single source within each cluster. Since all $\{\tau_{uv}\}$ are positive continuous random variables, the observed timestamps are distinct with probability one. The following definitions are illustrated in \cref{fig:obs_adm}.

\begin{Definition}\label{def: ad_set}
A node $v_k\in \calV$ is called \emph{observable} w.r.t.\ $u\in V$ if the following condition holds: for each pair of distinct nodes $v_i,v_j\in \calV\cap [u,v_k]$ such that $v_i\in [u,v_j)$, then $t_i< t_j$. The collection of observable nodes w.r.t.\ $u$ is called its observation cluster, and denoted by $\calV_u$. 
\end{Definition}

\begin{Definition}\label{def: ad_nodes}
Let $u\in V$. The collection of all nodes $v\in V$ such that $\calV_v = \calV_u$ is denoted by $A_u$, which is called the source candidate cluster w.r.t.\ $\calV_u$.
\end{Definition}

The observation cluster $\calV_u$ of $u$ is the set of observed infected nodes in $\calV$ that can possibly be infected by $u$. However, $\calV_u$ may also be the observation cluster of other nodes $v\ne u$. We collect all these potential source nodes into the source candidate cluster $A_u$. From \cref{def: ad_set,def: ad_nodes}, we have the following immediate observations. For a $v_i\in\calV$, since $v_i\in \calV_{v_i}$, the set $\calV_{v_i}$ is non-empty. Similarly, $v_i\in A_{v_i}$. Let $\calI_{s_m}$ be the collection of nodes in $\calV$ infected by $s_m\in S$. Specifically, if there is only one source $s$, then $\calV_s=\calI_{s}=\calV$. For any $u\in V$ , let 
\begin{align}\label{nu}
\nu_u \triangleq \argmin_{v_k\in \calV_u} t_k.
\end{align}
We have the following results regarding $\calV_u$, $A_u$, and $\calI_{s_m}$.

\begin{Lemma} \label{lem:aica2}
	Consider any node $u\in V$. 
	\begin{enumerate}[(i)]
		\item \label[claim]{L.1} If $v_k\in \calV_u$, then for any $v_i\in[u,v_k)\cap \calV$, we have $v_i\in \calV_u$ and $t_i< t_k$. 
		\item \label[claim]{L.2} $A_u\cap \calV=\{\nu_u\}$. Specifically, if $u=\nu_{s_m}$ for some $s_m\in S$, then $\calV_u=\calV_{s_m}$ (or $s_m\in A_u$).
		\item \label[claim]{L.3} $A_u$ is connected. 
		\item \label[claim]{L.4} For each $s_m\in S$, $\calI_{s_m}\subset \calV_{s_m}$.
		\item \label[claim]{L.5} If $v_k\in \calV_u$, then $\calV_{v_k}\subset \calV_u$. Specifically, if $v_k\in \calI_{s_m}$, then $\calV_{v_k}\subset \calV_{s_m}$.
		\item \label[claim]{L.6} If $\nu_u\in \calV_v$, then $\calV_u\subset \calV_v$.
	\end{enumerate}
\end{Lemma}

\begin{IEEEproof}\ 
	\begin{enumerate}[(i)]
		\item This follows directly from \cref{def: ad_set}.
		\item We first prove that for any $v_k\in\calV\backslash\nu_u$ then $\calV_u\neq \calV_{v_k}$. If $v_k\notin \calV_u$, then it follows directly that $\calV_u\neq \calV_{v_k}$. If $v_k\in \calV_u\backslash \nu_u$, then $\nu_u\notin \calV_{v_k}$ because $t_k>t'$, where $t'$ is the timestamp of node ${\nu_u}$. Since $\nu_u\in \calV_{u}$, we have $\calV_u\neq \calV_{v_k}$. We then prove that $\calV_u=\calV_{\nu_u}$. If $u\in \calV$, then $u=\nu_u$ and it follows directly that $\calV_u=\calV_{\nu_u}$. We next consider the case where $u\notin \calV$. Suppose $v_k\in \calV_u\backslash \nu_u$. According to \cref{L.1}, we find $v_i\in\calV_u$ such that $v_i\in[u,v_k]$ and $[u,v_i]\cap\calV=v_i$. By the minimality condition \cref{nu}, $[u,\nu_u]\cap \calV=\{\nu_u\}$. Then we have $[\nu_u,v_i]\cap \calV=\{\nu_u,v_i\}$ and $t'< t_i$, thus $v_i\in\calV_{\nu_u}$; and moreover $v_k\in\calV_{\nu_u}$ from \cref{def: ad_set}. Suppose on the contrary $v_k\notin \calV_{u}$. Then from \cref{def: ad_set}, there exists a pair of distinct nodes $v_i,v_j\in \calV\cap [u,v_k]$ such that $v_i\in [u,v_j)$. As a consequence, $t_i> t_j$. Since $[u,\nu_u]\cap \calV=\{\nu_u\}$, we have $v_i\in [\nu_u,v_j)$; and hence $v_k\notin \calV_{\nu_u}$. Therefore $\calV_u=\calV_{\nu_u}$.
		\item According to \cref{L.2}, it is equivalent to prove $A_{\nu_u}$ is connected. We now prove that for any $v_k\in \calV$, $A_{v_k}$ is connected. From \cref{L.2} we have $A_{v_k}\cap \calV=\{v_k\}$. Therefore given $u\in A_{v_k}\backslash v_k$, we have $u\notin \calV$. We show that $(v_k,u)\cap\calV=\emptyset$. Suppose on the contrary that there exists $v_i\in \calV$ and $v_i\in (v_k,u)$. According to \cref{L.1}, since $v_k\in \calV_u$ then $v_i\in\calV_u$ and $t_i<t_k$. Then we must have $v_i\notin\calV_{v_k}$ which contradicts $\calV_u=\calV_{v_k}$. Therefore for any $v\in(v_k,u)$ we have $[u,v]\cap\calV=\emptyset$, then it is easy to see that $\calV_u=\calV_v=\calV_{v_k}$ thus $v\in A_{v_k}$. Therefore the claim of connectedness holds.
		\item If $v_k\in \calI_{s_m}$, then we have $[s_m,v_k]\cap \calV\subset \calI_{s_m}$. As the propagation delay over each edge is positive, it is easy to verify that $v_k\in \calV_{s_m}$ from Definition \ref{def: ad_set}. 
		\item We prove that if $v_i\in \calV_u$ and $v_j\in \calV_{v_i}$, then $v_j\in \calV_u$. From Definition \ref{def: ad_set}, for each pair $v_a,v_b\in [u,v_i]\cap \calV$ such that $v_a\in [u,v_b]$, we have $t_a\leq t_b\leq t_i$. If $v_j\in \calV_{v_i}$ then for each pair $v_c,v_d\in [v_i,v_j]\cap \calV$ such that $v_c\in [v_i,v_d]$, we have $t_i\leq t_c\leq t_d$. So we have $[u,v_i]\cap [v_i,v_j]\cap \calV=v_i$, and then for each pair $v_p,v_q\in [u,v_j]\cap \calV$ such that $v_p\in [u,v_q]$, we have $t_p\leq t_q$. This implies that $v_j\in \calV_u$. Specifically, if $v_k\in \calI_{s_m}$, then $v_k\in \calV_{s_m}$ thus $\calV_{v_k}\subset \calV_{s_m}$. 		
		\item According to \cref{L.2} and \cref{L.5}, if $w=\nu_u\in \calV_v$, then $\calV_u=\calV_w\subset \calV_v$. Specifically, if $\nu_{s_i}\in \calV_{s_j}$, then $\calV_{s_i}\subset \calV_{s_j}$.   
	\end{enumerate} 
\end{IEEEproof}

From \cref{lem:aica2}\ref{L.4}, we see that the set of observed nodes $\calI_{s_m}$ infected by $s_m$, is a subset of $\calV_{s_m}$. Therefore $\calV_{s_m}$ can be used as an approximation of $\calI_{s_m}$. We provide a procedure to approximate $\{\calV_{s_m}: s_m \in S\}$ in the following.

For a given $v_i\in\calV$, our aim is to develop a method to efficiently generate $A_{v_i}$. Starting from the root $v_i$, we perform BFS, with the search path stopping immediately before hitting any node in $\calV\backslash \{v_i\}$. 
Let $B(v_i,\calV)$ be the collection of discovered nodes in this tree. 

\begin{Proposition} \label{prop:AB}
	For any $v_i\in\calV$, $A_{v_i}\subset B(v_i,\calV)$. If $v_i\notin \calV_{v_j}$ for any $v_j\in\calV\backslash\{v_i\}$, then $A_{v_i}=B(v_i,\calV)$.
\end{Proposition}

\begin{IEEEproof}
	According to the proof of \cref{lem:aica2}\ref{L.3}, it is easy to see that if $u\in A_{v_i}$, then $u\in B(v_i,\calV)$. Given the condition $v_i\notin \calV_{v_j}$ for any $v_j\in\calV\backslash{v_i}$, we prove that if $u\in B(v_i,\calV)$ then $u\in\calV_{v_i}$. According to \cref{lem:aica2}\ref{L.2}, it is equivalent to prove $\nu_u=v_i$. Suppose on the contrary $\nu_u=v_k$ and $v_k\neq v_i$. Then $[u,v_k]\cap\calV=v_k$. Since $u\in B(v_i,\calV)$ then $[u,v_i]\cap\calV=v_i$ and $v_i\in\calV_u$. So we have $[v_i,v_k]\cap\calV=\{v_i,v_k\}$ and $t_k<t_i$. Therefore $v_i\in\calV_{v_k}$; and we obtain a contradiction. Then we must have $\nu_u=v_i$ and $u\in\calV_{v_i}$, which completes the proof.    
\end{IEEEproof}

We now introduce the notion of an admissible covering of $\calV$.
\begin{Definition}
A covering that decomposes $\calV = \bigcup_{i=1}^l \calU_i$ into a union (not necessarily disjoint) of non-empty subsets $\calU_i$ for $1\leq i\leq l$, is called \emph{admissible} if for each $i$, and there exists $u_i$ such that $\calV_{u_i}=\calU_i$. 
\end{Definition}

It is easy to see that $\{\calV_{s_m}\}_{m=1}^{|S|}$ is an admissible covering of $\calV$. However, since the sources are unknown, we propose the Multiple Sources Reduction (MSR) algorithm in Algorithm~\ref{Alg_MSR} to find an admissible covering. We have the following result.

\begin{algorithm}[!htb]
	\caption{Multiple Sources Reduction (MSR)}
	\label{Alg_MSR}
	\begin{algorithmic}[1]		
		\REQUIRE Adjacency matrix of the tree $G$, observations $\calV=\{v_1,\ldots,v_n\}$ and $\bT=[t_1,\ldots,t_n]'$.
		\ENSURE $\{\calV_{\xi_i}\}_{i=1}^l$ and $\{A_{\xi_i}\}^l_{i=1}$. 
		\STATE Let $\calU=\calV$, and $i=1$.
		\WHILE {$\calU\neq\emptyset$}
		\STATE $\xi_i= \argmin_{v_k\in\calU} t_k$.
		\STATE Find $\calV_{\xi_i}$ according to \cref{def: ad_set}, then find $A_{\xi_i}=B(\xi_i,\calV)$ (cf.\ \cref{Prop: covering}\ref{it:A_nu}).
		\STATE Set $l=i$, $i=i+1$, $\calU=\calU\backslash \calV_{\xi_i}$. 
		\ENDWHILE
	\end{algorithmic}
\end{algorithm}

\begin{Proposition} \label{Prop: covering} 
	Consider the the MSR algorithm in Algorithm~\ref{Alg_MSR}. 
	\begin{enumerate}[(i)]
		\item\label[claim]{it:A_nu} For each $i=1,\ldots,l$, $A_{\xi_i}=B(\xi_i,\calV)$.
		\item $\{\calV_{\xi_i}\}_{i=1}^l$ is an admissible covering of $\calV$. 
		\item \label[claim]{P.2} Each $A_{\xi_i}$ contains at least one source, so $l\leq |S|$. If $l=|S|$, then each $A_{\xi_i}$ contains exactly one source.
		\item \label[claim]{P.3} $l=|S|$ if and only if for each distinct pair $s_i,s_j\in S$, $\nu_{s_i}\notin \calV_{s_j}$. 
	\end{enumerate}
\end{Proposition}

\begin{figure}[!htb]
	\footnotesize	
	\centering
	\subfigure{		
		\begin{minipage}[b]{.4\linewidth}
			\centering
			\includegraphics[height=1.5cm]{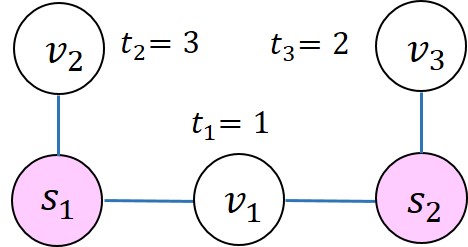}
			\centerline{(a)}
		\end{minipage}%
	}%
	\subfigure{
		\begin{minipage}[b]{.6\linewidth}
			\centering
			\includegraphics[height=1.5cm]{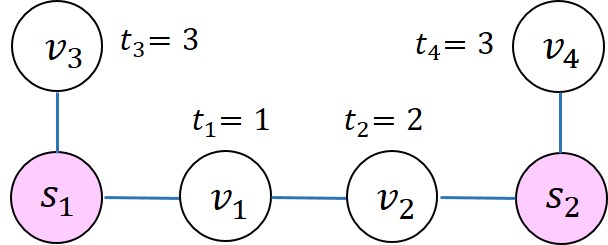}
			\centerline{(b)}
		\end{minipage}
	}
	\caption{Two cases where $l<|S|$. The observed infected nodes are labeled with their infection timestamps, and $|S|=2$. Since $\xi_1=v_1=\nu_{s_1}$ and $\calV_{\xi_1}=\calV$ then we have $l=1 < |S|=2$ and $\nu_{s_2}\in \calV_{s_1}$. In (a), $A_{\xi_1}=\{s_1,s_2,v_1\}$. In (b), $A_{\xi_1}=\{s_1,v_1\}$.}
	\label{fig:two_cases}
\end{figure}

\begin{IEEEproof}\
	\begin{enumerate}[(i)]
		\item Since $\xi_i= \argmin_{v_k\in\calU} t_k$, for any $v_j\in\calU\backslash\{\xi_i\}$, we have $\xi_i\notin\calV_{v_j}$. Suppose there exists $v_j\in\calV\backslash\calU$ such that $\xi_i\in\calV_{v_j}$. Since $v_j\in\calV_{\xi_k}$ for some $k<i$, from \cref{lem:aica2}\ref{L.5}, $\calV_{v_j}\subset\calV_{\xi_k}$ and we obtain a contradiction with $\xi_i\notin \calU$. The claim then follows from \cref{prop:AB}.
		\item It follows directly that each $\calV_{\xi_i}$ is non-empty and $\calV=\bigcup_{i=1}^l \calV_{\xi_i}$ with $l\leq |\calV|$.
		\item If $\xi_1\in \calV_{s_1}$, then $\xi_1= \nu_{s_1}$. According to Lemma \ref{lem:aica2}\ref{L.2}, $s_1\in A_{\xi_1}$. If $\calU\neq \emptyset$ and suppose $\xi_i\in \calV_{s_i}$, then from Lemma \ref{lem:aica2}\ref{L.6}, we must have $\nu_{s_i} \notin \bigcup_{j=1}^{i-1} \calV_{\xi_j}$. Therefore $\nu_{s_i}=\xi_i$ and $s_i\in A_{\xi_i}$. This procedure is then repeated and each $A_{\xi_i}$ contains at least one source. Moreover, it is likely that some sources are not contained in any $A_{\xi_i}$. Refer to the example in \figref{fig:two_cases}. So $l\leq |S|$. If $l=|S|$, then each $A_{\xi_i}$ contains exactly one source. 
		\item By the proof of \cref{P.2}, if for each distinct pair  $s_i,s_j\in S$ such that $\nu_{s_i}\notin \calV_{s_j}$, then we can find $\{\xi_m\}_{m=1}^{|S|}$ with $\xi_m=\nu_{s_m}$. On the contrary, if there exists $\nu_{s_i}\in \calV_{s_j}$ then  $\nu_{s_j}< \nu_{s_i}$, so we must choose $\nu_{s_j}$ before $\nu_{s_i}$. From Lemma \ref{lem:aica2}\ref{L.6}, $\calV_{s_i}\subset \calV_{s_j}$ thus $\calV_{s_i}\cap \calU\backslash \calV_{s_j}=\emptyset$. Hence $s_i\in A_{\nu_{s_j}}$ (cf.\ \figref{fig:two_cases}(a)) or $s_i$ will not be contained at any $\{A_{\xi_k}\}^l_{k=1}$ (cf.\ \figref{fig:two_cases}(b)). This implies that $l < |S|$.     
	\end{enumerate}
\end{IEEEproof}

Finding $\xi_i$ in \cref{Alg_MSR} can be accomplished by maintaining a min-heap for $\bT$, which incurs an overall time complexity of $O(|\calV|)$ for the whole algorithm. Finding $A_{\xi_i}$ involves BFS, which incurs a time complexity of $O(|V \backslash\cup_{k<i} A_{\xi_k}|)$. Therefore, the overall time complexity of MSR is~$O(|V|)$. 

\Cref{Prop: covering}\ref{P.2} indicates that if we assume that there is only a single source in each source candidate cluster $A_{\xi_i}$, then we are underestimating the number of sources (see \cref{fig:two_cases} for an example). Therefore we adopt an additional procedure developed in \cite{Ji2016} to further partition each $A_{\xi_i}$ and $\calV_{\xi_i}$. The paper \cite{Ji2016} has proposed an algorithm, which we call the Source-Set Size Estimation (SSSE) algorithm, to estimate the number of infection sources in a tree by partitioning the tree into several covering components. We briefly summarize the SSSE algorithm in \cref{Alg_SSSE}.

\begin{algorithm}[!htb]
	\caption{Source-Set Size Estimation (SSSE)\cite{Ji2016}}
	\label{Alg_SSSE}
	\begin{algorithmic}[1]		
		\REQUIRE Adjacency matrix of the tree $T$.
		\ENSURE A set of disjoint and connected subtrees of $T$, denoted by $\{T_i\}_{i=1}^K$.
		\STATE Compute the average pairwise distance $\bar{d}$ of $T$, and find a path $P=[u,v]$ with the maximal length.
		\IF {$|P|>\bar{d}$} 
		\STATE Define a function $s_P(\cdot)$ on $P$: For each $w\in P$, find the leaves (with degree 1) in $T$ that are closest to $w$ compared with other nodes in $P\backslash w$. Then $s_P(w)$ is defined to be the sum of the distances between $w$ and each of these leaves. If $s_P(w)$ is a local minimum and $w'$ is a neighbor of $w$ in $P$, then delete the edge connecting $w$ and $w'$. $T$ is divided into two subtrees $T_1$ and $T_2$. 
		\ENDIF 
		\STATE For $T_1$ and $T_2$, repeat the above partition procedure until the diameters of the subtrees are less than $\bar{d}$ or $s_P(\cdot)$ does not have a local minimum.
	\end{algorithmic}
\end{algorithm}

The time complexity of SSSE is $O(|V|)$. From \cref{lem:aica2}\ref{L.3} we see that each $A_{\xi_i}$ forms a connected subtree. So we apply the SSSE algorithm on each $A_{\xi_i}$ to obtain a new covering for $\calV$, which are denoted by $\{C_i\}_{i=1}^L$, where $L \geq l$. At the same time $\{\calV_{\xi_i}\}_{i=1}^l$ is also partitioned as $\{O_i\}_{i=1}^L$. Our simulations indicate that it is now possible for $L>|S|$ as SSSE may over-estimate the number of sources in each $A_{\xi_i}$. 

Finally, for each $1\leq i\leq L$, we apply GSSI to find a source estimate $\hat{s}_i\in C_i$ using $O_i$ as an approximation for $\calI_{s_i}$. Note that different sources may have different diffusion parameters $(t_0, \mu, \sigma^2)$ and GSSI is able to estimate these parameters for each possible source. An example to illustrate the steps involved in the estimation of multiple sources on a tree is given in \cref{fig:obs_adm}.

\subsection{Multiple Sources Estimation for General Graphs}\label{SCCE}

We now heuristically extend the sources estimation procedure for trees described in the previous subsection to general graphs, and summarized it in \cref{Alg_SCCE}, where $\{\hat{s}_i\}^L_{i=1}$ is the set of estimated sources. First, we make use of the BFS heuristic in generalizing the MSR algorithm to a general graph before applying SSSE. Notice that for general graphs, it is possible that before applying the SSSE algorithm, the number of source candidate clusters $l>|S|$ and $A_{\xi_i}\cap S=\emptyset$ for some values of $i$. The generalized MSR and SSSE steps involve a time complexity of $O(|V|^2)$ since we find BFS trees rooted at each $\xi_i$. For each subgraph we run GSSI, whose time complexity depends on the number of nodes and observed nodes of the subgraph. The worst case complexity is incurred when $L=1$. Therefore the overall time complexity of SCCE is $O(|V|^3 + |V||\calV|^3)$.

\begin{algorithm}[!htb]
	\caption{Source Candidate Clustering and Estimation (SCCE)}
	\label{Alg_SCCE}
	\begin{algorithmic}[1]		
		\REQUIRE Adjacency matrix of the graph $G$, observation $\calV=\{v_1,\ldots,v_n\}$ and $\bT=[t_1,\ldots,t_n]'$.
		\ENSURE $\hat{S}=\{\hat{s}_i\}^L_{i=1}$
		\STATE Let $\calU=\calV$, and $i=1$
		\WHILE {$\calU\neq\emptyset$}
		\STATE $\xi_i= \argmin_{v_k\in\calU} t_k$.
		\STATE Find one BFS tree rooted at $\xi_i$ and then obtain $B(\xi_i,\calV)$ and $\calV_{\xi_i}$ on this tree. Set $A_{\xi_i}=B(\xi_i,\calV)$. 
		\STATE Set $l=i$, $i=i+1$, $\calU=\calU\backslash \calV_{\xi_i}$.
		\ENDWHILE
		\STATE For each $1\leq i\leq l$, run SSSE algorithm to partition $\calV_{\xi_i}$ and $A_{\xi_i}$ on the BFS tree rooted at $\xi_i$. Finally we obtain $\{C_i\}_{i=1}^L$ and $\{O_i\}_{i=1}^L$.
		\FOR {every $C_i$ and $O_i$}
        \STATE Run GSSI algorithm. Compute $f_s$ in \cref{GSSI:fs} of \cref{Alg_GSSI} for each $s\in C_i\backslash\{\xi_i\}$. 
		\STATE Pick $\hat{s}_i$ to be a node $s$ that minimizes $f_s$.
		\ENDFOR
	\end{algorithmic}
\end{algorithm} 

\subsection{Experimental Evaluations for Multiple Sources Estimation}
\label{sect:experiments SCCE}

We evaluate the performance of our SCCE algorithm with the same graphs used in \cref{subsect:exp_GSSI}. We compare SCCE with the method proposed in \cite{fu2016multi}, which we call the BDSL algorithm. To quantify the performance of SCCE and BDSL, we adopt the performance metric (with a slight modification for fairer comparison) proposed in \cite{Luo2013}: We first match the estimated sources $\{\hat{s}_i\}_{i=1}^L$ with the actual sources $\{{s}_i\}_{i=1}^{|S|}$ so that the sum of the error distances between each estimated source and its match is minimized. Denote this matching as $\pi$. If we incorrectly estimate the number of infection sources, i.e., $L\neq |S|$, we add a penalty $\eta ||S|-L|$. Denote $d_{\text{max}}$ and $\overline d$ as the diameter and average pairwise distance of the graph. Since the distance between a source and a random node does not exceed $d_{\text{max}}$, a reasonable choice of penalty $\eta\leq d_{\text{max}}$. We then define the average error to be
\begin{align}\label{metirc: Delta}
\Delta=\frac{1}{\calL}\left(\sum_{i=1}^{\min(L,|S|)}d(s_i,\hat{s}_{\pi(i)})+\eta||S|-L|\right),
\end{align}
where $d(u,v)$ is the length of the shortest path in $G$ between $u$ and $v$, and
\begin{align*}
\calL=\left\{ 
\begin{array}{ll}
\min(L,|S|)\quad &\text{if }\eta=0,\\
\max(L,|S|)\quad &\text{if }0<\eta\leq d_{\text{max}}. 
\end{array} \right.
\end{align*}

For different applications, we may assign different values to $\eta$ depending on how important it is to estimate correctly the number of infection sources. In this paper, we consider $\eta=0,\overline d$ and the extreme case where $\eta=d_{\text{max}}$. The BDSL algorithm cannot estimate the number of sources, so we assume it has prior knowledge of $|S|$. Therefore for BDSL, we have $\calL=L=|S|$.

\begin{figure}[!tb]
	\centering 
	\subfigure[]{
		\begin{minipage}[b]{0.48\linewidth}
			\centering
			\includegraphics[height=3.5cm]{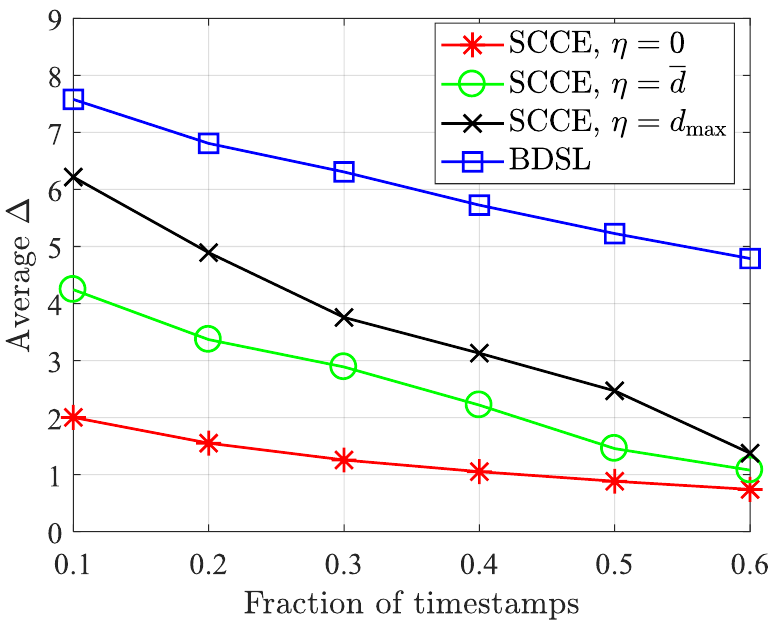}
		\end{minipage}
		\label{fig:SCCE_AB_2_dist}
	}%
	\subfigure[]{
		\begin{minipage}[b]{0.48\linewidth}
			\centering
			\includegraphics[height=3.5cm]{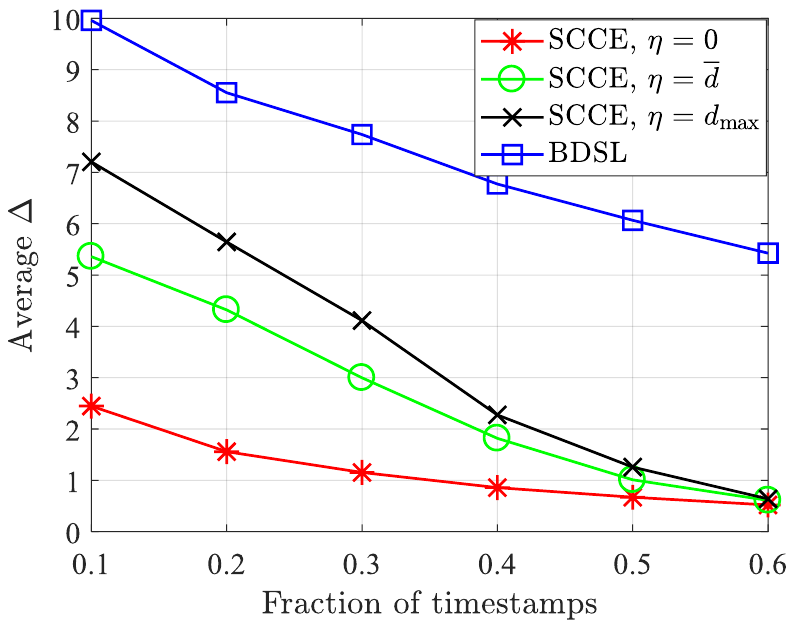}
		\end{minipage}
		\label{fig:SCCE_ER_2_dist}
	}
	\subfigure[]{
		\begin{minipage}[b]{0.8\linewidth}
			\centering
			\includegraphics[height=3.5cm]{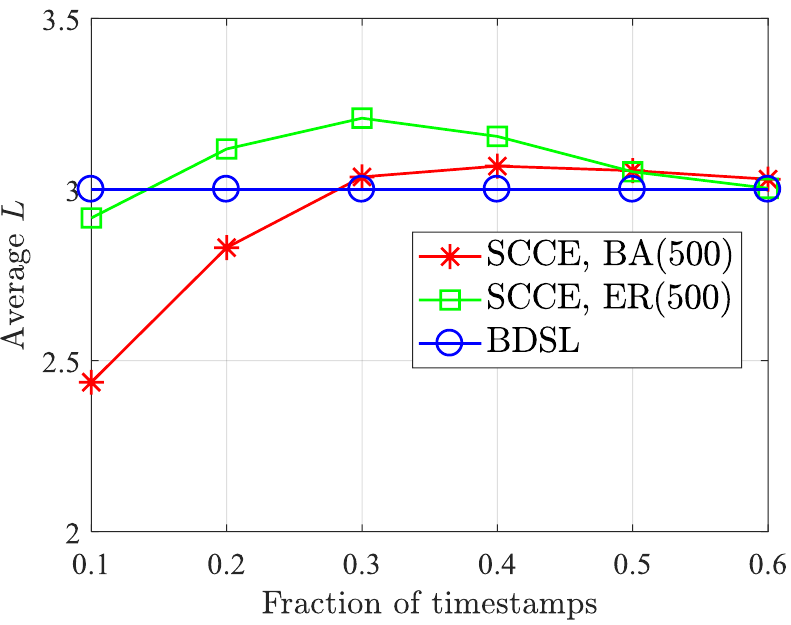}
		\end{minipage}
		\label{fig:SCCE_2_size}
	}
	\caption{Multiple sources estimation on (a) $\BA(500)$ and (b) $\ER(500)$.} \label{fig:SCCE_2}
\end{figure}

We first perform simulations on $\ER(500)$ and $\BA(500)$. For each simulation, we let the number of sources be uniformly chosen from $\{2,3,4\}$. From \cref{Prop: covering}\ref{P.3}, we see that if two sources are close to each other, then it is likely that one of them cannot be identified. So we choose sources such that the distance between each pair of sources is no less than the average pairwise distance of the tree, which is a reasonable assumption in real applications.\footnote{Note that no estimation method can find all the sources correctly with probability one. Therefore, in real applications, source estimation methods are used to narrow down the potential candidate sources so that further investigation on them and their neighbors within a small radius can be conducted. Sources that are close to each other can then be considered as a single source under our estimation framework.}  For the diffusion process, let the start times of all sources be identical, which is an assumption required by BDSL, and propagation delays along all edges follow a truncated Gaussian distribution $\calN (2,1)$. Notice that BDSL requires to know the mean of the Gaussian distribution used while our algorithm does not. For the BDSL algorithm, since we uniformly choose $|S|$ from $\{2,3,4\}$, the average $L$ is 3, which is used as a benchmark to compare how well SCCE performs in estimating the number of sources. We randomly choose a portion of nodes as observed nodes for the multiple sources estimation. We average $\Delta$ in \cref{metirc: Delta}, and $L$ over 300 simulations and plot against the fraction of timestamps as shown in \cref{fig:SCCE_2}. The plots in \cref{fig:SCCE_AB_2_dist,fig:SCCE_ER_2_dist} show that SCCE performs better than BDSL for both B-A and E-R trees although SCCE requires less information. For SCCE algorithm, from \cref{fig:SCCE_2_size} we see that $L>|S|$ in some cases due to the introduction of the additional partition procedure developed in the SSSE algorithm; and the average $L$ is close to the average $|S|$ if we observe enough timestamps. 

\begin{figure}[!htb]
	\centering 
		\begin{minipage}[b]{0.5\linewidth}
			\centering
			\includegraphics[height=3.5cm]{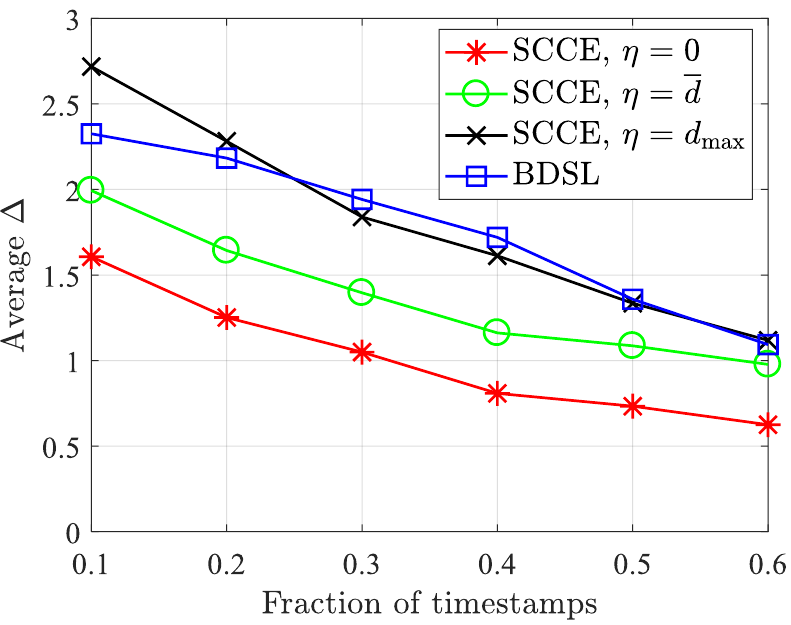}
			\centerline{(a)}
		\end{minipage}%
		\begin{minipage}[b]{0.5\linewidth}
			\centering
			\includegraphics[height=3.5cm]{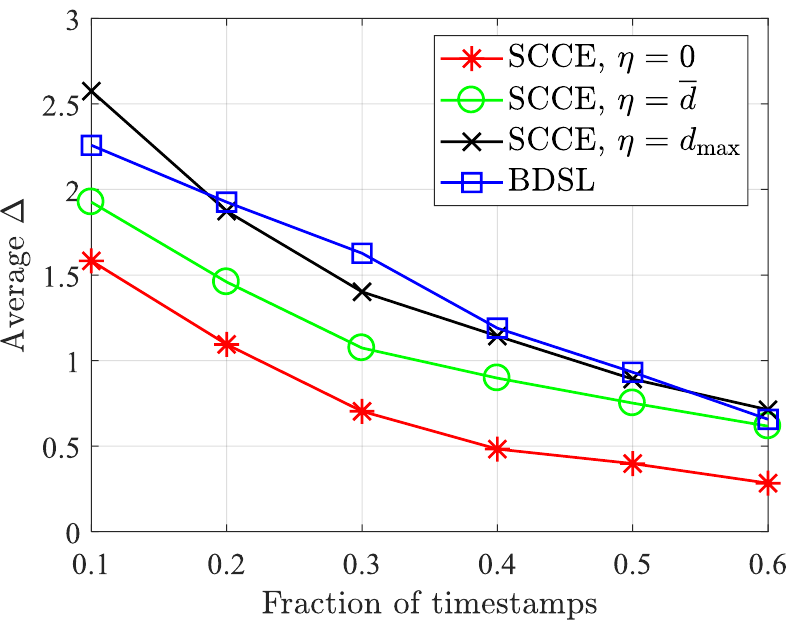}
			\centerline{(b)}
		\end{minipage}
	\caption{Multiple sources estimation on (a) $\BA(500,16)$ and (b) $\ER(500,16)$.} \label{fig:SCCE_16}
\end{figure}

\begin{figure}[!htb]
	\centering 
		\begin{minipage}[b]{0.5\linewidth}
			\centering
			\includegraphics[height=3.53cm]{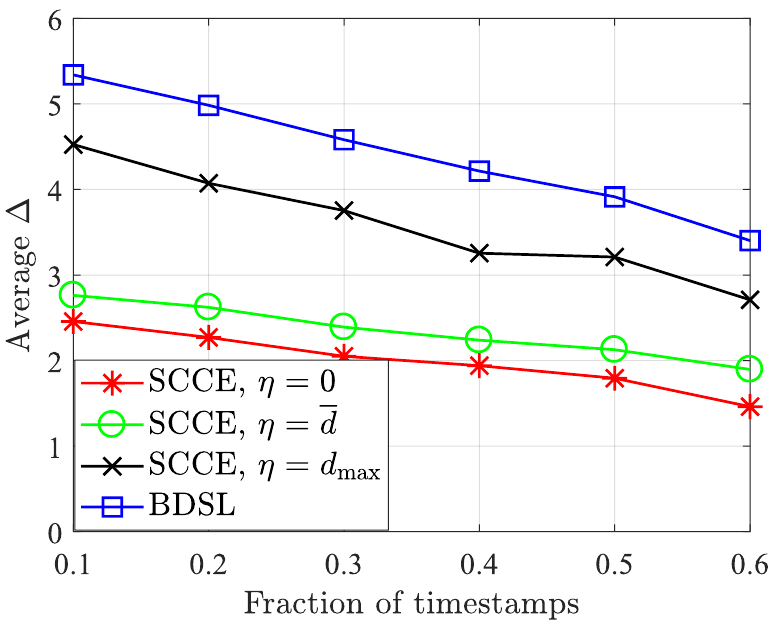}
			\centerline{(a)}
		\end{minipage}%
		\begin{minipage}[b]{0.5\linewidth}
			\centering
			\includegraphics[height=3.53cm]{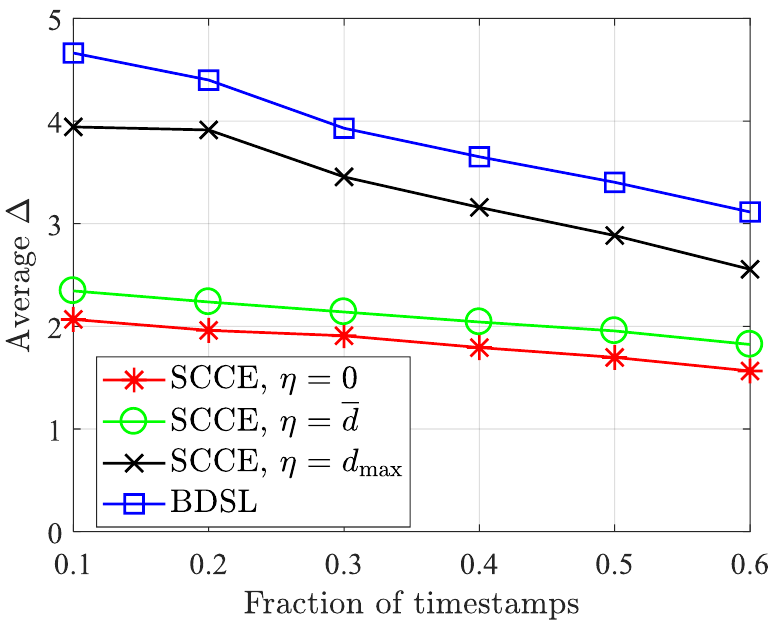}
			\centerline{(b)}
		\end{minipage}
	\caption{Multiple sources estimation on (a) Enron network and (b) Facebook network.} \label{fig:SCCE_RN}
\end{figure}

\begin{figure}[!htb]
	\centering 
		\begin{minipage}[b]{0.5\linewidth}
			\centering
			\includegraphics[height=3.5cm]{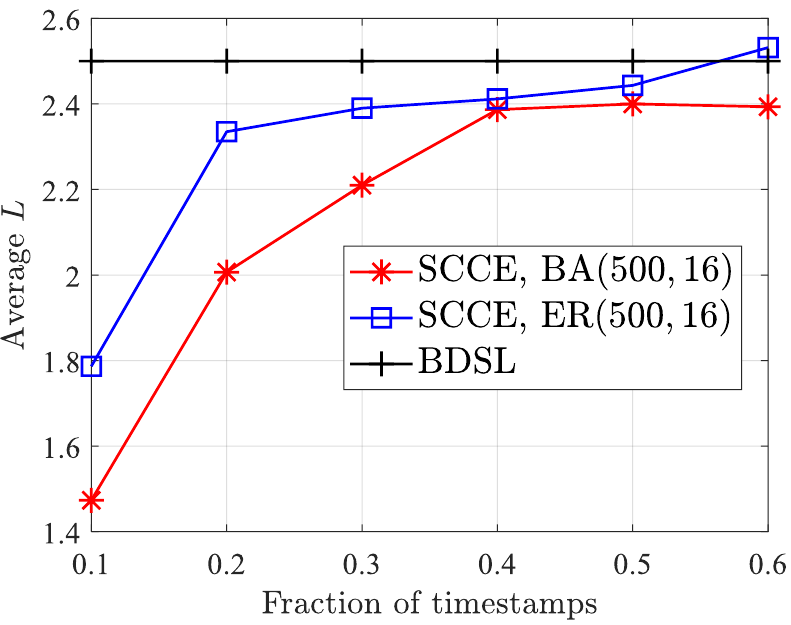}
		\end{minipage}%
		\begin{minipage}[b]{0.5\linewidth}
			\centering
			\includegraphics[height=3.5cm]{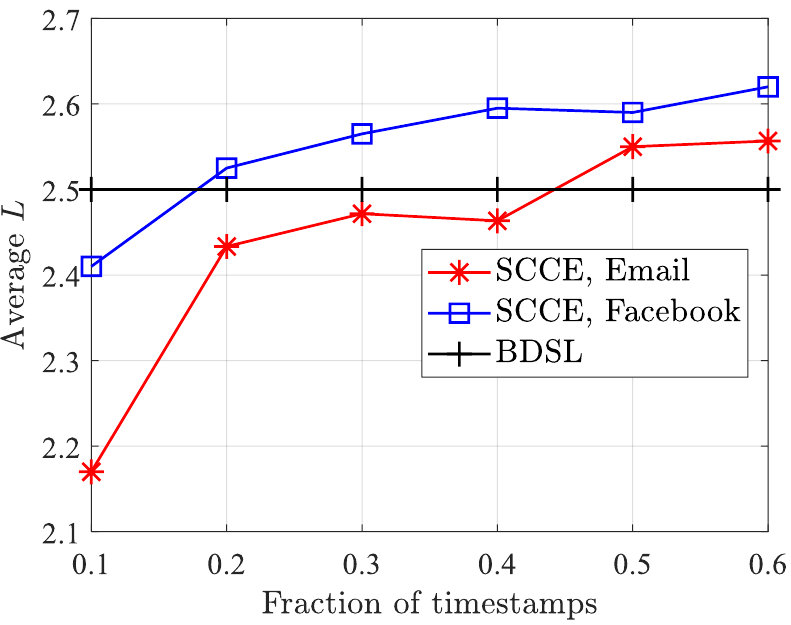}
		\end{minipage}
	\caption{Performance of the average $L$ for different graphs.}
    \label{fig:SCCE_size}
\end{figure}

We next perform simulations on some synthetic graphs and real networks (cf.\ \cref{tab:graph_properties} for some properties of these graphs). Let the number of sources be uniformly chosen from $\{2,3\}$, so that its average value is 2.5. For the truncated Gaussian distribution used to simulate the infection spreading, we set $(\mu, \sigma^2)=(3,1)$. Since in real applications, $|S|$ is typically not very large, we make the additional assumption that $|S|\leq5$ when running the SCCE algorithm by controlling the number of iterations in the MSR step. Simulation results in \cref{fig:SCCE_16,fig:SCCE_RN,fig:SCCE_size} indicate that in terms of average $\Delta$, SCCE performs best {in the cases where $\eta=0$ and $\overline{d}$. In most cases SCCE also performs better than BDSL even when $\eta=d_{\text{max}}$.} For the performance of average $L$, we see that generally SCCE finds more source estimates as the fraction of timestamps increases. Sometimes we may over-estimate the number of sources. On E-R and B-A graphs, SCCE is able to find estimates that are on average within 2 hops of the real sources when the fraction of timestamps is at least $20\%$. For the Enron and Facebook networks, BDSL has poor estimation accuracy, while in the case of $\eta\leq\overline{d}$, SCCE produce estimates that are on average within 2.5 hops of the real sources with no less than $30\%$ of nodes being observed. 

\vspace{-1em}
\vspace{-1em}

\begin{figure}[!htb]
	\centering 
	\begin{minipage}[b]{0.5\linewidth}
		\centering
		\includegraphics[height=3.5cm]{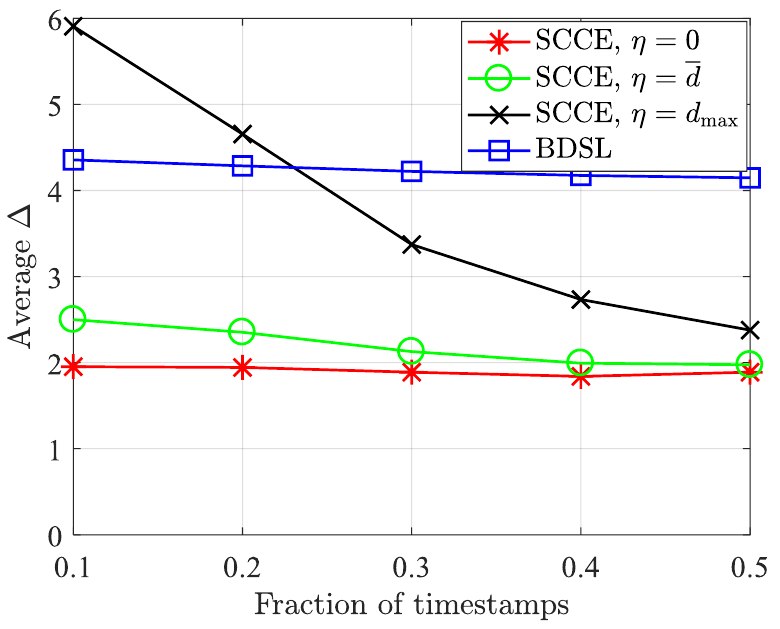}
	\end{minipage}%
	\begin{minipage}[b]{0.5\linewidth}
		\centering
		\includegraphics[height=3.5cm]{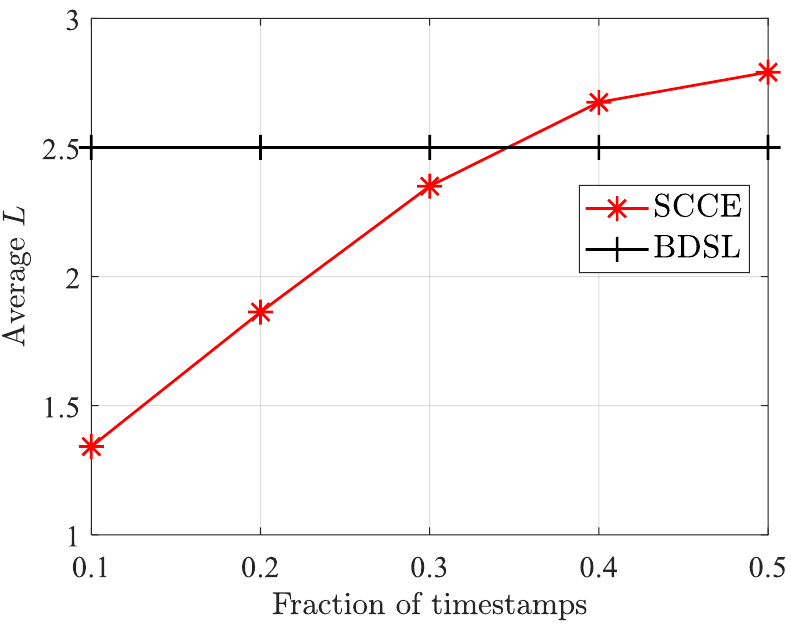}
	\end{minipage}
	\caption{Malware sources estimation on Facebook network.}
	\label{fig:malware_SCCE}
\end{figure}

We also apply our algorithm to estimate multiple sources in malware propagation using the model and parameters discussed in \cref{subsect:exp_GSSI}. We randomly and uniformly choose 2 or 3 malware sources and collect infection timestamps. Simulation results are shown in \cref{fig:malware_SCCE}. The plots show that BDSL performs poorly even with increasing fraction of timestamps; we believe this is because the estimation of the mean propagation delays are not accurate. For SCCE, if the penalty $\eta$ dose not exceed $\overline{d}$, we are able to estimate sources that have an average error of 2.5 even with only $10\%$ timestamps. The performance does not improve much as the fraction of timestamps increases. We believe this is because the diffusion model we have assumed does not match the characteristics of malware propagation.
  
Finally, we evaluate the performance of SCCE with a sample of real data provided by SNAP.\footnote{https://snap.stanford.edu/data/higgs-twitter.html} First developed in the work \cite{de2013}, the Higgs dataset has been built after monitoring the spreading processes on Twitter before, during and after the announcement of the discovery of a new particle with the features of the elusive Higgs boson on 4 July 2012. Before the announcement on 1 July 2012, there were some rumors about the discovery of a Higgs-like boson at Tevatron. Such rumors were posted on the Twitter social network and quickly spread among Twitter users around the world. The dataset provides a social network with 456,626 users and 14,855,842 connections. In addition, a retweet network including the users and timestamps of retweet events is also given. We extract a connected sub-network with 5,078 users and 26,449 connections. All the users in this sub-network posted or retweeted the rumors related to the Higgs boson discovery. We also collect 4,254 users' first retweet timestamps and the total spanned time is about 39 hours starting from 1 July 2012. We then model the sub-network as an undirected graph (cf.\ Table \ref{tab:graph_properties}), regard the timestamps as our observations, and the first several users who posted the rumors independently as infection sources. 

The diffusion of the retweets is heterogeneous and unknown to us, which means that our assumptions for the diffusion model are not satisfied. However, we still apply our SCCE algorithm to estimate the sources given the graph topology and timestamp observations. Algorithms such as GAU and BDSL require more information such as the mean of the propagation delays, which need to be learned from historical data. From the dataset we find three infection sources, let $S=\{s_1,s_2,s_3\}$. In the extracted graph we have $d(s_1,s_2)=2$, $d(s_2,s_3)=2$, and $d(s_1,s_3)=3$. These three infection sources have different start times. We observe that $s_1$ first initiated the diffusion process, then after about 7 and 15 hours, $s_2$ and $s_3$ started their diffusion, respectively. We then randomly choose a fraction of timestamps to perform multiple sources estimation using SCCE. After obtaining the estimated sources, we compute the average $\Delta$ and average $L$ to show its performance. To test the BDSL algorithm, we learn the mean time of a tweet spreading across an edge from 50,000 historical tweets, which yields an estimate of about 4 hours. We average over 50 experiments. The comparison between SCCE and BDSL is shown in \cref{tab:real_data}. We see that SCCE performs well on this real data. Comparatively, BDSL has poorer accuracy since the start times of the sources differ significantly from each other, and the mean propagation delays are not accurately known. The average number of estimated sources $L$ obtained by SCCE is 2.2 with 10\% timestamps, 3.1 with 20\% timestamps and 3.3 with 30\% timestamps.   

\begin{table}[!htbp]
	\centering  
	\begin{tabular}{|l|c|c|c|c|}  
		\hline
		\multirow{3}{*}{\makecell*[l]{Fraction of \\timestamps\\  }} & \multicolumn{4}{c|}{Average $\Delta$}\\ \cline{2-5} 
     	&SCCE   &SCCE  &SCCE  &\multirow{2}{*}{\makecell*[c]{BDSL}} \\   
     	& ($\eta=0$)&($\eta=\overline{d}$) &($\eta=d_{\text{max}}$) &    \\ \hline
		\multicolumn{1}{|l|}{$10\%$} &0.73 &1.88 &3.2 &5.6  \\ \hline       
		\multicolumn{1}{|l|}{$20\%$} &1.10 &1.72 &1.92 &5.3 \\ \hline  
		\multicolumn{1}{|l|}{$30\%$} &0.96 &1.68 &1.89 &5.4 \\ 
		\hline
	\end{tabular}\\
	\caption{Comparison between SCCE and BDSL on real data.}
	\label{tab:real_data}
\end{table}

To summarize, with the same fraction of timestamps, the performance of multiple sources estimation is not as good as single source estimation. This makes sense as the number of timestamps used for each source estimation is reduced. A summary comparison of SCCE (with $\eta=\overline{d}$ and $d_{\text{max}}$) with BDSL when 30\% timestamps are observed is provided in \cref{tab:SCCE_compare}.

\begin{table}[!htbp]
	\centering  
	\begin{tabular}{|l|c|c|}  
		\hline
		\multirow{2}{*}{\makecell*[l]{Graph }} & \multicolumn{2}{c|}{Error reduction(\%)}\\\cline{2-3}
		&$\eta=\overline{d}$   &$\eta=d_{\text{max}}$\\
		\hline   
		$\ER(500)$ & 61.2 &46.9 \\       
		$\BA(500)$ & 54.4 &40.4 \\   
		$\ER(500,16)$ &34.4 &14.1 \\ 
		$\BA(500,16)$ &27.8 &5.2 \\  
		Enron & 47.8 &18.1\\   
		Facebook & 45.6 &12.0 \\ 
		Facebook (Malware)& 49.5&43.8\\
		Twitter &68.9&65.0\\
		\hline
	\end{tabular}
	\caption{Percent of reduced average error distance of SCCE compared with BDSL.}
	\label{tab:SCCE_compare}
\end{table}

\section{Conclusion}\label{sect:conclusion}
We have developed algorithms to estimate infection sources using a subset of timestamps for the cases where there is a single source, and where there are multiple sources and the number of sources is unknown. We adopt a Gaussian spreading model with unknown diffusion parameters, which are estimated as part of our inference algorithms. In the single source case, we introduce a new heuristic that optimizes over a parametrized family of Gromov matrices to more accurately identify the source in a general graph. In the multiple sources estimation problem, we develop theory that allows us to partition the observed infected nodes into observation clusters. We then apply our single source algorithm in each cluster. Experimental evaluations with synthetic and real-world data suggest that our approaches can find the infection sources to within a small number of hops from the true sources with a small number of timestamp observations.

The works \cite{Fanti2015, Luo2015} have developed infection strategies that make it difficult for a network administrator to estimate the infection source accurately. These works assume that the network administrator has access only to the infection status of each node, but not the infection timestamps. With infection timestamps, it becomes more difficult to obfuscate the identity of the source. Since our approach involves estimation of the diffusion parameters and can handle model mismatch to some extent, as shown in our simulations on malware source estimation, we expect our approach to be somewhat robust to obfuscation strategies. Designing good source obfuscation strategies when timestamp information is available, and corresponding source estimation strategies to counter them are potential interesting future research directions.

\appendices

\section{Heuristic Analysis of GSSI}\label[appendix]{app:heuristic}

Suppose that the estimates by GSSI $(\hat{s},[\hat{t}_0,\hat{\mu}]',\hat{\sigma}^2) = (s_1, \bbeta_0, \sigma^2)$, the true parameter values. Then, $\alpha_{\hat{s}}=\alpha_{s_1}$ in \cref{eq:alpha_s} is the solution of the following optimization problem:
\begin{align*}
\alpha_{s_1}
&=\argmax_{\alpha\in [0,1]}{p(\bT\mid s_1,t_0,\mu,\sigma,\alpha)} \\
&=\argmax_{\alpha\in [0,1]}{\frac{\exp\left(-\frac{1}{2\sigma^2}\bU'\bA_{s_1}^{-1}(\alpha)\bU \right)}{\sqrt{(2\pi\sigma^2)^n\det \bA_{s_1}(\alpha)}}}\nonumber\\
&=\argmin_{\alpha\in [0,1]} f(\alpha)\triangleq\log\det \bA_{s_1}(\alpha)+\frac{1}{\sigma^2}\bU'\bA_{s_1}^{-1}(\alpha)\bU, 
\end{align*}
where $\bU=\bT-\bD_{s_1}\bbeta_0$ and $\bA_{s_1}(\alpha)=\alpha \bH_{s_1}+(1-\alpha)\bLambda_{s_1}$ according to \cref{eq:linear_comba}. Let $f_1(\alpha)=\log \det \bA_{s_1}(\alpha)$ and $f_2(\alpha)=\bU'\bA_{s_1}^{-1}(\alpha)\bU$ so that $f(\alpha)=f_1(\alpha)+f_2(\alpha)/\sigma^2$. Our goal is to show that on average, $f(0)$ is close to a local optimum when $G$ is a tree, since $\alpha_{\hat{s}}=0$ makes GSSI equivalent to MLE.

\begin{Lemma_A} \label{lem:alpha_opt}
Suppose that $G$ is a tree, and the GSSI estimates $(\hat{s},[\hat{t}_0,\hat{\mu}]',\hat{\sigma}^2) = (s_1, \bbeta_0, \sigma^2)$. Then, we have
	\begin{enumerate}[(i)]
	\item \label[claim]{L1.1} $f_1(\alpha)$ is concave and non-decreasing for $\alpha\in[0,1]$. 
	\item \label[claim]{L1.2}$f_2(\alpha)$ is convex w.r.t.\ $\alpha$ and $\frac{d}{d\alpha}f_2(\alpha)\Big|_{\alpha=0}\geq\bU'\bLambda_{s_1}^{-1}\bU(1-\lambda_{\max}(\bLambda_{s_1}^{-1}\bH_{s_1}))$, where $\lambda_{\max}(\cdot)$ is the largest eigenvalue of its matrix argument.
	\item\label[claim]{L1.3}$\E[\frac{d}{d\alpha}f(\alpha)\Big|_{\alpha=0}]=0$.
	\end{enumerate}
\end{Lemma_A}
\begin{IEEEproof}\
	\begin{enumerate}[(i)]
	\item  The first and second derivatives of $f_1(\alpha)$ w.r.t.\ $\alpha$ are given by
	\begin{align}
	\frac{d}{d\alpha}f_1(\alpha)&=\frac{1}{\det \bA_{s_1}(\alpha)}\frac{d}{d\alpha}\det \bA_{s_1}(\alpha)\nonumber\\
	&=\tr(\bA_{s_1}^{-1}(\alpha)\frac{d}{d\alpha}\bA_{s_1}(\alpha))\nonumber\\
	&=\tr(\bA_{s_1}^{-1}(\alpha)(\bH_{s_1}-\bLambda_{s_1})),\label{d_f1}
	\end{align}
	and
	\begin{align*}
		\ddfrac{^2}{\alpha^2}f_1(\alpha)&=\tr(-\bA_{s_1}^{-1}(\alpha)\frac{d}{d\alpha}\bA_{s_1}(\alpha)\bA_{s_1}^{-1}(\alpha)\\
		&\quad\cdot(\bH_{s_1}-\bLambda_{s_1}))\\
	&=\tr(-\bA_{s_1}^{-1}(\alpha)(\bH_{s_1}-\bLambda_{s_1})\bA_{s_1}^{-1}(\alpha)\\
	&\quad\cdot(\bH_{s_1}-\bLambda_{s_1}))\\
	&\leq 0. 
	\end{align*}
	Hence $f_1(\alpha)$ is concave. We also have
	\begin{align*}
	\frac{d}{d\alpha}f_1(\alpha)\Bigr|_{\alpha=1}&=\tr(\bH_{s_1}^{-1}(\bH_{s_1}-\bLambda_{s_1}))\\
	&=n-\tr(\bH_{s_1}^{-1}\bLambda_{s_1})\\
	&=n-\sum_{i=1}^{n}[\bLambda_{s_1}]_{i,i}/[\bH_{s_1}]_{i,i}\\
	&=0,
	\end{align*}
	therefore $\frac{d}{d\alpha}f_1(\alpha)\geq 0$ for all $\alpha\in[0,1]$.
	\item We have
	\begin{align}
	\frac{d}{d\alpha}f_2(\alpha)&=-\bU'\bA_{s_1}^{-1}(\alpha)\frac{d}{d\alpha}\bA_{s_1}(\alpha)\bA_{s_1}^{-1}(\alpha)\bU\nonumber\\
	&=-\bU'\bA_{s_1}^{-1}(\alpha)(\bH_{s_1}-\bLambda_{s_1})\bA_{s_1}^{-1}(\alpha)\bU,\label{d_f2}
	\end{align}
	and
	\begin{align*}
	\frac{d^2f_2}{d\alpha^2}(\alpha)&=2\bU'\bA_{s_1}^{-1}(\alpha)\frac{d}{d\alpha}\bA_{s_1}(\alpha)\bA_{s_1}^{-1}(\alpha)\\
	&\quad\cdot(\bH_{s_1}-\bLambda_{s_1})\bA_{s_1}^{-1}(\alpha)\bU\\
	&=2[\bU'\bA_{s_1}^{-1}(\alpha)(\bH_{s_1}-\bLambda_{s_1})]\bA_{s_1}^{-1}(\alpha)\\
	&\quad\cdot[(\bH_{s_1}-\bLambda_{s_1})\bA_{s_1}^{-1}(\alpha)\bU]\\
	&\geq 0,
	\end{align*}
	where the last inequality holds because $\bA_{s_1}(\alpha)$ is positive semidefinite. Therefore, $f_2(\alpha)$ is convex.
	
	To show the lower bound, we have
	\begin{align*}
	\frac{d}{d\alpha}f_2(\alpha)\Bigr|_{\alpha=0}&=\bU'\bLambda_{s_1}^{-1}\bU-\bU'\bLambda_{s_1}^{-1}\bH_{s_1}\bLambda_{s_1}^{-1}\bU.
	\end{align*}
	Let $\bW=\bH_{s_1}^{\frac{1}{2}}\bLambda_{s_1}^{-\frac{1}{2}}$ and $\bu=\bLambda_{s_1}^{-\frac{1}{2}}\bU$. Let $\sigma_{\max}(\bW)$ be the spectral norm  of $\bW$, then we obtain from Theorem 5.2.7 of \cite{meyer2000},
	\begin{align*}
	\sigma_{\max}^2(\bW)&=\sup_{\bx\neq \bm{0}}\frac{\bx'\bW'\bW\bx}{\bx'\bx}\\
	&\geq\frac{\bu'\bW'\bW\bu}{\bu'\bu}\\
	&=\frac{\bU'\bLambda_{s_1}^{-1}\bH_{s_1}\bLambda_{s_1}^{-1}\bU}{\bU'\bLambda_{s_1}^{-1}\bU},
	\end{align*}
	and $\bU'\bLambda_{s_1}^{-1}\bH_{s_1}\bLambda_{s_1}^{-1}\bU\leq\sigma_{\max}^2(\bH_{s_1}^{\frac{1}{2}}\bLambda_{s_1}^{-\frac{1}{2}}) \bU'\bLambda_{s_1}^{-1}\bU$. Since $\sigma_{\max}^2(\bW)=\lambda_{\max}(\bW\bW')$, we have
	\begin{align*}
	\frac{d}{d\alpha}f_2(\alpha)\Bigr|_{\alpha=0}&\geq \bU'\bLambda_{s_1}^{-1}\bU(1-\lambda_{\max}(\bLambda_{s_1}^{-1}\bH_{s_1})),
	\end{align*}
	since $\bH_{s_1}^{\ofrac{2}}\bLambda_{s_1}^{-1}\bH_{s_1}^{\ofrac{2}}$ and $\bLambda_{s_1}^{-1}\bH_{s_1}$ have the same eigenvalues.
		
	\item From \cref{eq:ml_pro}, we have $\bT\sim\calN(\bD_{s_1}\bbeta_0,\sigma^2\bLambda_{s_1})$ so $\frac{1}{\sigma}\bLambda_{s_1}^{-\frac{1}{2}}\bU\sim\calN(0,\bI_n)$, and $\frac{1}{\sigma^2}\bU'\bLambda_{s_1}^{-1}\bU$ follows a chi-squared distribution with $n$ degrees of freedom. From \cref{d_f1,d_f2}, we obtain
	\begin{align*}
	\E[\frac{d}{d\alpha}f(\alpha)\Bigr|_{\alpha=0}]=&\tr(\bLambda_{s_1}^{-1}(\bH_{s_1}-\bLambda_{s_1}))\\
	&+\frac{1}{\sigma^2}\E[\bU'\bLambda_{s_1}^{-1}\bU]\\
	&-\frac{1}{\sigma^2}\E[\bU'\bLambda_{s_1}^{-1}\bH_{s_1}\bLambda_{s_1}^{-1}\bU]\\
	=&\tr(\bLambda_{s_1}^{-1}\bH_{s_1})-n+n\\
	&-\tr(\bLambda_{s_1}^{-\frac{1}{2}}\bH_{s_1}\bLambda_{s_1}^{-\frac{1}{2}})\\
	=&0,
	\end{align*}
	and the proof is complete.
	\end{enumerate}
\end{IEEEproof}

\Cref{lem:alpha_opt}\ref{L1.3} says that on average the first derivative of $f(\alpha)$ is zero at $\alpha=0$, which implies that $\alpha_{s_1}$ cannot be much larger than zero with high probability.

\bibliographystyle{IEEEtran}
\bibliography{IEEEabrv,reference}
\end{document}

%% file: preamble.tex
\usepackage{setspace} 
\usepackage[T1]{fontenc}
\usepackage{amsmath,amssymb,amsfonts,mathrsfs,bm}
\usepackage{amsthm}
\usepackage{cite}
\usepackage{array}
\usepackage[shortlabels]{enumitem}
\usepackage{graphicx}
\usepackage{url}
\usepackage{color}
\usepackage{algorithm,algorithmic}
\usepackage{multirow}
\usepackage{breqn}
\usepackage[table]{xcolor}
\usepackage[normalem]{ulem}
\usepackage{array}
\newcolumntype{L}[1]{>{\raggedright\let\newline\\\arraybackslash\hspace{0pt}}m{#1}}
\newcolumntype{C}[1]{>{\centering\let\newline\\\arraybackslash\hspace{0pt}}m{#1}}
\newcolumntype{R}[1]{>{\raggedleft\let\newline\\\arraybackslash\hspace{0pt}}m{#1}}
\usepackage{subfigure}

\usepackage{xparse}

\ifx\notloadhyperref\undefined
	\ifx\loadbibentry\undefined
		\usepackage[hidelinks]{hyperref} 
	\else
		\usepackage{bibentry}
		\makeatletter\let\saved@bibitem\@bibitem\makeatother
		\usepackage[hidelinks]{hyperref}
		\makeatletter\let\@bibitem\saved@bibitem\makeatother
	\fi
\else
\relax
\fi

\usepackage[capitalize]{cleveref}
\crefname{equation}{}{}
\Crefname{equation}{}{}
\crefname{claim}{claim}{claims}
\crefname{step}{step}{steps}
\crefname{line}{line}{lines}
\crefname{Theorem}{Theorem}{Theorems}
\crefname{Corollary}{Corollary}{Corollaries}
\crefname{Proposition}{Proposition}{Propositions}
\crefname{Lemma}{Lemma}{Lemmas}
\crefname{Definition}{Definition}{Definitions}
\crefname{Example}{Example}{Examples}
\crefname{Assumption}{Assumption}{Assumptions}
\crefname{Remark}{Remark}{Remarks}
\crefname{Theorem_A}{Theorem}{Theorems}
\crefname{Corollary_A}{Corollary}{Corollaries}
\crefname{Proposition_A}{Proposition}{Propositions}
\crefname{Lemma_A}{Lemma}{Lemmas}
\crefname{Definition_A}{Definition}{Definitions}

\ifx\useTheoremCounter\undefined
\newtheorem{Theorem}{Theorem}
\newtheorem{Corollary}{Corollary}
\newtheorem{Proposition}{Proposition}
\newtheorem{Lemma}{Lemma}
\else

\newtheorem{Proposition}[theorem]{Proposition}
\fi

\newtheorem{Definition}{Definition}

\newtheorem{Lemma_A}{Lemma}[section]

\theoremstyle{remark}

\newcommand{\Real}{\mathbb{R}}

\newcommand{\calI}{\mathcal{I}}

\newcommand{\calL}{\mathcal{L}}

\newcommand{\calN}{\mathcal{N}}

\newcommand{\calT}{\mathcal{T}}
\newcommand{\calU}{\mathcal{U}}
\newcommand{\calV}{\mathcal{V}}

\newcommand{\bA}{\mathbf{A}}

\newcommand{\bD}{\mathbf{D}}

\newcommand{\bH}{\mathbf{H}}

\newcommand{\bI}{\mathbf{I}}

\newcommand{\bM}{\mathbf{M}}

\newcommand{\bT}{\mathbf{T}}
\newcommand{\bu}{\mathbf{u}}
\newcommand{\bU}{\mathbf{U}}

\newcommand{\bW}{\mathbf{W}}
\newcommand{\bx}{\mathbf{x}}

\DeclareSymbolFont{bsfletters}{OT1}{cmss}{bx}{n}
\DeclareSymbolFont{ssfletters}{OT1}{cmss}{m}{n}
\DeclareMathSymbol{\bsfGamma}{0}{bsfletters}{'000}
\DeclareMathSymbol{\ssfGamma}{0}{ssfletters}{'000}
\DeclareMathSymbol{\bsfDelta}{0}{bsfletters}{'001}
\DeclareMathSymbol{\ssfDelta}{0}{ssfletters}{'001}
\DeclareMathSymbol{\bsfTheta}{0}{bsfletters}{'002}
\DeclareMathSymbol{\ssfTheta}{0}{ssfletters}{'002}
\DeclareMathSymbol{\bsfLambda}{0}{bsfletters}{'003}
\DeclareMathSymbol{\ssfLambda}{0}{ssfletters}{'003}
\DeclareMathSymbol{\bsfXi}{0}{bsfletters}{'004}
\DeclareMathSymbol{\ssfXi}{0}{ssfletters}{'004}
\DeclareMathSymbol{\bsfPi}{0}{bsfletters}{'005}
\DeclareMathSymbol{\ssfPi}{0}{ssfletters}{'005}
\DeclareMathSymbol{\bsfSigma}{0}{bsfletters}{'006}
\DeclareMathSymbol{\ssfSigma}{0}{ssfletters}{'006}
\DeclareMathSymbol{\bsfUpsilon}{0}{bsfletters}{'007}
\DeclareMathSymbol{\ssfUpsilon}{0}{ssfletters}{'007}
\DeclareMathSymbol{\bsfPhi}{0}{bsfletters}{'010}
\DeclareMathSymbol{\ssfPhi}{0}{ssfletters}{'010}
\DeclareMathSymbol{\bsfPsi}{0}{bsfletters}{'011}
\DeclareMathSymbol{\ssfPsi}{0}{ssfletters}{'011}
\DeclareMathSymbol{\bsfOmega}{0}{bsfletters}{'012}
\DeclareMathSymbol{\ssfOmega}{0}{ssfletters}{'012}

\newcommand{\bbeta}{\bm{\beta}}

\newcommand{\bLambda}{\bm{\Lambda}}

\DeclareMathOperator*{\argmax}{arg\,max}
\DeclareMathOperator*{\argmin}{arg\,min}

\DeclareMathOperator{\diag}{diag}

\DeclareMathOperator{\tr}{tr}

\newcommand{\qednew}{\nobreak \ifvmode \relax \else
      \ifdim\lastskip<1.5em \hskip-\lastskip
      \hskip1.5em plus0em minus0.5em \fi \nobreak
      \vrule height0.75em width0.5em depth0.25em\fi}

\newcommand{\ofrac}[1]{{\frac{1}{#1}}}
\newcommand{\ddfrac}[2]{{\frac{d {#1}}{d {#2}}}}

\newcommand{\cond}[2]{\left. {#1}\, \middle| \, {#2} \right.}

\DeclareDocumentCommand \P { g d() g } {%
	\IfNoValueTF {#3} 
	{%
		\IfNoValueTF {#1} 
		{%
			\IfNoValueTF {#2}
			{%
				\mathbb{P}%
			}%
			{%
				\mathbb{P}\left({#2}\right)%
			}%
		}%
		{%
			\IfNoValueTF {#2}
			{%
				\mathbb{P}_{#1}%
			}%
			{%
				\mathbb{P}_{#1}\left({#2}\right)%
			}%
		}%
	}%
	{%
		\IfNoValueTF {#1} 
		{%
			\mathbb{P}\left(\cond{#2}{#3}\right)%
		}%
		{%
			\mathbb{P}_{#1}\left(\cond{#2}{#3}\right)%
		}%
	}%
}

\DeclareDocumentCommand \E { g o g } {%
	\IfNoValueTF {#3} 
	{%
		\IfNoValueTF {#1} 
		{%
			\IfNoValueTF {#2}
			{%
				\mathbb{E}%
			}%
			{%
				\mathbb{E}\left[{#2}\right]%
			}%
		}%
		{%
			\IfNoValueTF {#2}
			{%
				\mathbb{E}_{#1}%
			}%
			{%
				\mathbb{E}_{#1}\left[{#2}\right]%
			}%
		}%
	}%
	{%
		\IfNoValueTF {#1} 
		{%
			\mathbb{E}\left[\cond{#2}{#3}\right]%
		}%
		{%
			\mathbb{E}_{#1}\left[\cond{#2}{#3}\right]%
		}%
	}%
}

\definecolor{gray90}{gray}{0.9}

\newcommand{\msout}[1]{\text{\color{green} \sout{\ensuremath{#1}}}}
\newcommand{\del}[1]{{\color{green}\ifmmode \msout{#1}\else\sout{#1}\fi}}

\newcommand{\hide}[1]{}

\newcommand{\figref}[1]{\figurename~\ref{#1}}
\renewcommand{\figurename}{Fig.}
\graphicspath{{./Figures/}} 